\documentclass[11pt]{article}
\usepackage[utf8]{inputenc}

\usepackage{amsmath}
\usepackage{graphicx, psfrag, epsf}
\usepackage{enumerate}  
\usepackage{url} 
\usepackage{multirow}
\usepackage{tabularx}
\usepackage{comment}

\usepackage{algorithm,algpseudocode}

\algnewcommand\algorithmicinput{\textbf{Input:}}
\algnewcommand\algorithmicoutput{\textbf{Output:}}
\algnewcommand\Input{\item[\algorithmicinput]}%
\algnewcommand\Output{\item[\algorithmicoutput]}%

\usepackage[margin=1in]{geometry}

\usepackage{setspace}

\usepackage{amssymb, amsfonts, amsthm, bm}
\usepackage{mathtools, mathrsfs,amsmath}
\usepackage{relsize} 
\usepackage{hyperref}
\usepackage{float}
\usepackage{enumitem}
\usepackage{enumerate}
\usepackage{verbatim}
\usepackage{wrapfig}
\usepackage{float}
\usepackage{booktabs}
\usepackage{comment}

\newtheorem{theorem}{Theorem}
\newtheorem{lemma}{Lemma}

\newtheorem{assumption}{Assumption}

\DeclareMathOperator*{\argmax}{argmax} 

\usepackage{color}

\title{Scalable marginalization of correlated latent variables  with applications to learning particle interaction kernels} 
\author{Mengyang Gu$^ \dag $\footnote{Correspondence should be addressed to Mengyang Gu (\href{mailto:mengyang@pstat.ucsb.edu}{mengyang@pstat.ucsb.edu} )}\, , Xubo Liu$^ \dag  $, Xinyi Fang$^ \dag  $, Sui Tang$^\ddag  $\\ \\
           \small $^ \dag  $ Department of Statistics and Applied Probability, University of California, Santa Barbara, CA \\
                      \small $^\ddag  $ Department of Mathematics, University of California, Santa Barbara, CA 
}
\date{}

\begin{document}

\maketitle


\begin{abstract}
Marginalization of latent variables or nuisance parameters is a fundamental aspect of Bayesian inference and uncertainty quantification. In this work, we focus on scalable marginalization of latent variables in modeling correlated data, such as  spatio-temporal or functional observations. We first introduce Gaussian processes (GPs) for modeling correlated data and highlight the computational challenge, where the computational complexity increases cubically fast along with the number of observations. We then review the  connection between the state space model and GPs with Mat{\'e}rn covariance for temporal inputs. The Kalman filter and Rauch-Tung-Striebel smoother were introduced as a scalable marginalization technique for computing the likelihood and making predictions of GPs without approximation. We summarize   recent efforts on extending the scalable marginalization idea to the linear model of coregionalization for multivariate correlated output and spatio-temporal observations. In the final part of this work, we introduce a novel marginalization technique to estimate interaction kernels and forecast particle trajectories. The computational progress lies in the sparse representation of the inverse covariance matrix of the latent variables,
then applying conjugate gradient for  improving predictive accuracy with large data sets. The  computational advances achieved in this work outline a wide range of applications in  molecular dynamic  simulation, cellular migration, and agent-based models.

~~


\end{abstract}

\section{Introduction}
\label{sec:intro}

Given a set of latent variables in a model, do we fit a model with a particular set of latent variables, or do we  integrate out the latent variables when making predictions?  Marginalization of latent variables is an iconic feature of the Bayesian analysis. The art of marginalization in statistics can at least  be  traced back  to the De Finetti's theorem \cite{de1937prevision}, which  states that an infinite sequence $\{X_i\}^{\infty}_{i=1}$ is exchangeable,  if and if only if there exists a random variable $\theta \in \Theta$ with probability distribution $\pi(\cdot)$, and a conditional distribution $p(\cdot \mid \theta)$, such that
\begin{equation}
p(x_1,...,x_N) =\int \left\{\prod^N_{i=1}p(x_i\mid\theta) \right\}\pi(\theta)d\theta.
\label{equ:exchangeability}
\end{equation}
Marginalization of nuisance parameters for models with independent observations  has been comprehensively reviewed in \cite{berger1999integrated}.  Bayesian model selection \cite{berger1996intrinsic,barbieri2004optimal} and Bayesian model averaging  \cite{raftery1997bayesian}, as two other examples, both rely on the marginalization of parameters in each model. 

For spatially correlated data, the  Jefferys prior of the covariance parameters in a Gaussian process (GP),  which is proportional to the squared root of the Fisher information matrix of the likelihood, often leads to improper posteriors  \cite{berger2001objective}. The posterior of the covariance parameter becomes proper if the prior is derived based on the Fisher information matrix of the marginal likelihood, after marginalizing out the mean and variance parameters. The resulting prior, after marginalization, is a reference prior, which has been studied for modeling spatially correlated data and computer model emulation  \cite{paulo2005default,ren2012objective,kazianka2012objective,Gu2018robustness,mure2021propriety}.

Marginalization of latent variables has lately been aware  by the machine learning community as well, for purposes of uncertainty quantification and propagation. In  \cite{lakshminarayanan2017simple}, for instance, the deep ensembles of models with a scoring function were proposed to assess the uncertainty in deep neural networks, and it is closely related to Bayesian model averaging with a uniform prior on parameters. This approach was further studied in \cite{wilson2020bayesian}, where the importance of marginalization is highlighted. Neural networks with infinite depth were shown to be equivalent to a GP with a particular kernel function in \cite{neal2012bayesian}, and it was lately shown in \cite{lee2017deep} that the results of deep neural networks can be reproduced by GPs, where the latent nodes are marginalized out. 

In this work, we study the marginalization of latent variables for correlated data, particularly focusing on scalable computation. Gaussian processes have been ubiquitously used for modeling spatially correlated data \cite{banerjee2014hierarchical}   and emulating computer experiments \cite{sacks1989design}. Computing the likelihood in GPs and making predictions, however, cost $\mathcal O(N^3)$ operations, where $N$ is the number of observations, due to finding the inverse  and determinant of the covariance matrix. To overcome the computational bottleneck, various approximation approaches, such as inducing point approaches \cite{snelson2006sparse}, fixed rank approximation \cite{cressie2008fixed}, integrated nested Laplace approximation \cite{rue2009approximate}, stochastic partial differential equation representation \cite{lindgren2011explicit}, local Gaussian process approximation \cite{gramacy2015local}, and hierarchical nearest-neighbor Gaussian process models \cite{datta2016hierarchical},  circulant embedding \cite{stroud2017bayesian},  many of which can be summarized into the framework of Vecchia approximation \cite{vecchia1988estimation,katzfuss2021general}. 
Scalable computation of a GP model with a multi-dimensional input space and a smooth covariance function is of great interest in recent years. 

The \textit{exact} computation of GP models with smaller computational complexity was less studied in past. To fill this knowledge gap, we will first review the  stochastic differential equation  representation of a GP with the Mat{\'e}rn covariance and one-dimensional input variable \cite{whittle1963stochastic,hartikainen2010kalman}, where the solution can be written as a dynamic linear model \cite{west2006bayesian}.  Kalman filter  and Rauch–Tung–Striebel smoother \cite{kalman1960new,rauch1965maximum} can be implemented for computing the likelihood function and predictive distribution exactly, reducing the computational complexity of GP using a Mat{\'e}rn kernel with a half-integer roughness parameter and 1D input from $\mathcal O(N^3)$ to $\mathcal O(N)$ operations. Here, interestingly,  the latent states  of a GP model are  {marginalized} out in Kalman Filter iteratively. Thus the Kalman filter can be considered as an example of marginalization of latent variables, which leads to efficient computation. 
Note that the Kalman filter is not directly applicable for GP with multivariate inputs, yet GPs with some of the widely used covariance structures, such as the product or separable  kernel \cite{bayarri2007framework} and linear model of coregionalization \cite{banerjee2014hierarchical}, can be written as state space models on an augmented lattice \cite{gu2018generalized,10.1214/21-BA1295}. Based on this connection, we introduce a few extensions of scalable marginalization  for modeling incomplete matrices of correlated data. 

The contributions of this work are twofold. First, the computational scalability and efficiency of marginalizing latent variables for models of correlated data and functional data are less studied. Here we discuss the marginalization of latent states in the Kalman filter in computing the likelihood and making predictions, with only $\mathcal O(N)$ computational operations. We  discuss recent extensions on structured data with multi-dimensional input. Second, we develop new marginalization techniques to estimate interaction kernels of particles and to forecast trajectories of particles, which have wide applications in agent-based models \cite{couzin2005effective}, cellular migration \cite{henkes2011active}, and molecular dynamic simulation \cite{rapaport2004art}.   The computational gain comes from the sparse representation of inverse covariance of interaction kernels, and the use of the conjugate gradient algorithm \cite{hestenes1952methods} for iterative computation. Specifically, we reduce the computational order from $\mathcal O((nMDL)^3)+\mathcal  O(n^4L^2M^2D)$ operations in recent studies \cite{lu2019nonparametric,feng2021data} to $\mathcal  O(Tn^2MDL)+\mathcal O(n^2MDL \log(nMDL))$ operations   based on training data of $M$ simulation runs, each containing $n$ particles in a $D$ dimensional space at $L$ time points,  with $T$ being the number of iterations in the sparse conjugate gradient algorithm.  
This allows us to estimate interaction kernels of  dynamic systems with many more  observations. Here the sparsity comes from the use of the Mat{\'e}rn kernel, which is distinct from any of the approximation methods based on sparse covariance structures.

The rest of the paper is organized below. We first introduce the GP as a surrogate model for approximating computationally expensive simulations in Section \ref{sec:gp}. The state space model representation of a GP with Mat{\'e}rn covariance and temporal input is introduced in Section \ref{subsec:state_space}. We then review the Kalman filter as a computationally scalable technique to marginalize out latent states for computing the likelihood of a GP model and making predictions in Section \ref{subsec:KF}. In Section \ref{subsec:marginalization_multidim}, we discuss the extension of latent state marginalization in linear models of coregionaliztion for   multivariate functional data, spatial and spatio-temporal data on the incomplete lattice. The new computationally scalable algorithm for estimating interaction kernel and forecasting particle trajectories  is introduced in Section \ref{sec:particle_interactions}. 
We conclude this study and discuss a few potential research directions in Section \ref{sec:conlusion}. The code and data used in this paper are publicly available: \url{https://github.com/UncertaintyQuantification/scalable_marginalization}. 

\section{Background: Gaussian process}
\label{sec:gp}
We briefly introduce the GP model in this section. We focus on computer model emulation, where the GP emulator is often used as a surrogate model to approximate computer experiments  \cite{santner2003design}. Consider a real-valued unknown function $z(\cdot)$, modeled by a Gaussian stochastic process (GaSP) or Gaussian process (GP),  $z(\cdot) \sim  \mathcal{GP}(\mu(\cdot), \sigma^2K(\cdot, \cdot))$, meaning that, for any  inputs $\{ \mathbf x_1,\ldots,\mathbf x_N\}$  (with $\mathbf x_i$ being a $p\times 1$ vector), the marginal distribution of $\mathbf z=(z(\mathbf x_1),...,z(\mathbf x_N))^T$ follows a multivariate normal distribution,
\begin{equation}
  \mathbf z \mid \bm \beta, \,\sigma^2, \,{\bm \gamma} \sim \mathcal{MN} (\bm \mu, \sigma^2   {\mathbf R}  )\,,
 \label{equ:GP}
 \end{equation}
where $\bm \mu=(\mu(\mathbf x_1),..., \mu(\mathbf x_N))^T$ is a vector of mean or trend parameters, $\sigma^2$ is the unknown variance and $ {\mathbf  R}$ is the correlation matrix with the $(i,j)$ element modeled by a kernel $K \, (\mathbf x_i, \mathbf   x_j)$ with parameters $\bm \gamma$. It is common to model the mean by  $\mu(\mathbf x)=   \mathbf h(\mathbf x){\bm \beta} \,$, where $\mathbf h(\mathbf x)$ is a $1\times q$ row vector of basis function, and $\bm \beta$ is a $q\times 1$ vector of mean  parameters.  

When modeling spatially correlated data, the isotropic kernel is often used, where the input of the kernel only depends on the Euclidean distance $K(\mathbf x_a, \mathbf   x_b)=K(|| \mathbf x_a-\mathbf x_b|| )$. In comparison,
each coordinate of the latent function in computer experiments could have different physical meanings and units. Thus a product kernel is often used in constructing a GP emulator, such that correlation lengths can be different at each coordinate. 
For any  $\mathbf x_a=( x_{a1}, \ldots,  x_{ap})$ and $\mathbf x_b=( x_{b1}, \ldots,  x_{bp})$, the kernel function can be written as $K(\mathbf x_a, \mathbf x_b)=K_1( x_{a1}, x_{b1})\times...\times K_p( x_{ap}, x_{bp})$, where $K_l$ is a kernel for the $l$th coordinate with a distinct range parameter $\gamma_l$, for $l=1,...,p$. Some frequently used kernels $K_l$ include power exponential and Mat{\'e}rn kernel functions  \cite{rasmussen2006gaussian}. The Mat{\'e}rn kernel, for instance, follows
\begin{equation}
 K_l(d_l)=\frac{1}{2^{\nu_l-1}\Gamma(\nu_l)}\left(\frac{\sqrt{2\nu_l} d_l}{\gamma_l} \right)^{\nu_l} \mathcal K_{\nu_l} \left(\frac{\sqrt{2\nu_l} d_l}{\gamma_l} \right),
 \label{equ:matern}
 \end{equation}
 where $d_l=|x_{al}-x_{bl}|$,  $\Gamma(\cdot)$ is the gamma function, $\mathcal{K}_{\nu_l}(\cdot/\gamma_l)$ is the modified Bessel function of the second kind with the range parameter and roughness parameter  being $\gamma_l$ and $\nu_l$, respectively. The Mat{\'e}rn correlation has a closed-form expression when the roughness parameter is a half-integer, i.e. $\nu_l={2k_l+1}/{2}$ with $k_l\in \mathbb N$. It becomes the exponential correlation and Gaussian correlation, when $k_l=0$ and $k_l\to \infty$, respectively. The  GP with Mat{\'e}rn kernel is $\lfloor \nu_l-1 \rfloor$ mean square differentiable at coordinate $l$. This is a good property, as the differentiability of the process is directly controlled by the roughness parameter.

Denote mean basis of observations $\mathbf H=(\mathbf h^T(\mathbf x_1),...,\mathbf h^T(\mathbf x_N))^T$.  The parameters in GP contain mean parameters $\bm \beta$, variance parameter $\sigma^2$, and range parameters $\bm \gamma=(\gamma_1,...,\gamma_p)$. Integrating out the mean and variance parameters with respect to reference prior $\pi(\bm \beta, \sigma)\propto 1/\sigma^2$, the predictive distribution of any input $\mathbf x^*$ follows a student t distribution \cite{Gu2018robustness}:
 \begin{equation}
z({\mathbf x^{*}}) \mid \mathbf{z},\, \bm \gamma   \sim \mathcal T ( \hat z({\mathbf x}^{*}),\hat{\sigma}^2K^{**}, N - q)\,,
\label{equ:predictiongp}
\end{equation}
with $N-q$ degrees of freedom, where
\begin{align}
\hat{z} ({\mathbf x}^{*}) =& { \mathbf h({\mathbf x}^{*})} \hat{\bm{\beta}}+\mathbf{r}^T(\mathbf{x}^*){{\mathbf R}}^{-1}\left(\mathbf{z}-\mathbf H\hat{\bm{\beta}}\right), \label{equ:gppredmean}\\
\hat{\sigma}^2 =&(N-q)^{-1}{\left(\mathbf{z}-\mathbf H \hat{\bm{\beta}}\right)}^{T}{{\mathbf R}}^{-1}\left({\mathbf{z}}-\mathbf H\hat{\bm{\beta}}\right),  \\
 K^{**} =& K({\mathbf x^{*}}, {\mathbf x^{*}})-{ \mathbf{r}^T(\mathbf{x}^*){ {\mathbf R}}^{-1}\mathbf{r}(\mathbf{x}^*)} +  {\mathbf  h}^*(\mathbf x^*)^T \nonumber \\
&\times \left(\mathbf H^T{{\mathbf R}}^{-1}\mathbf H \right)^{-1} {\mathbf  h}^*(\mathbf x^*) , 
\end{align}
with $ {\mathbf h}^*(\mathbf x^*)=\left({{\mathbf h(\mathbf x^{*})}}-\mathbf H^T {{\mathbf R}}^{-1}\mathbf{r}(\mathbf{x}^*) \right)$, $\hat{\bm{\beta}}=\left( \mathbf H^T {{\mathbf R}}^{-1} \ \mathbf H \right)^{-1}\mathbf H^T{{\mathbf R}}^{-1}\mathbf{z}$ being the generalized least squares estimator of $\bm \beta$,    and $\mathbf{r}(\mathbf{x}^*) = (K(\mathbf{x}^*,{\mathbf{x}}_1 ), \ldots, K(\mathbf{x}^*,{\mathbf{x}}_N ))^T$.

Direct computation of the likelihood requires $\mathcal O(N^3)$ operations due to computing the Cholesky decomposition of the covariane matrix for matrix inversion, and the determinant of the covariance matrix. Thus a posterior sampling algorithm, such as the Markov chain Monte Carlo (MCMC) algorithm can be slow, as it requires a large number of posterior samples. 
Plug-in estimators, such as the maximum likelihood estimator (MLE) were often used to estimate the range parameters $\bm \gamma$ in covariance. In \cite{Gu2018robustness}, the maximum marginal posterior estimator (MMPE) with robust parameterizations was discussed to overcome the instability of the MLE. The MLE and MMPE of the parameters in a GP emulator with both the product kernel and the isotropic kernel are all implemented in the {\tt RobustGaSP} package \cite{gu2018robustgasp}.   

In some applications, we may not directly observe the latent function but a noisy realization: 
 \begin{equation}
 y(\mathbf x)= z(\mathbf x)+\epsilon(\mathbf x), 
 \label{equ:obs_noise}
 \end{equation} 
where $z(.)$ is modeled as a zero-mean GP with covariance $\sigma^2K(.,.)$, and $\epsilon(\mathbf x) \sim \mathcal N(0, \sigma^2_0)$ follows an independent Gaussian noise. This model is typically referred to as the Gaussian process regression \cite{rasmussen2006gaussian}, which is suitable for scenarios containing noisy observations, such as experimental or field observations,  numerical solutions of differential equations with non-negligible error, and stochastic simulations. 
Denote the noisy observations $\mathbf y=(y(\mathbf x_1),y(\mathbf x_2),...,y(\mathbf x_N))^T$ at the design input set $\{\mathbf x_1, \mathbf x_2,..., \mathbf x_N\}$ and the nugget parameter $\eta=\sigma^2_0/\sigma^2$. Both range  and nugget parameters in GPR can be estimated by the plug-in estimators \cite{gu2018robustgasp}. 
The predictive distribution of  $f(\mathbf x^*)$ at any input $\mathbf x^*$  can be obtained by replacing $\mathbf R$ with $\mathbf {\tilde R}=\mathbf R+\eta \mathbf I_n$ in Equation (\ref{equ:predictiongp}).

\begin{figure}[t] 
\centering
  \begin{tabular}{ccc}
 
    \includegraphics[width=.33\textwidth]{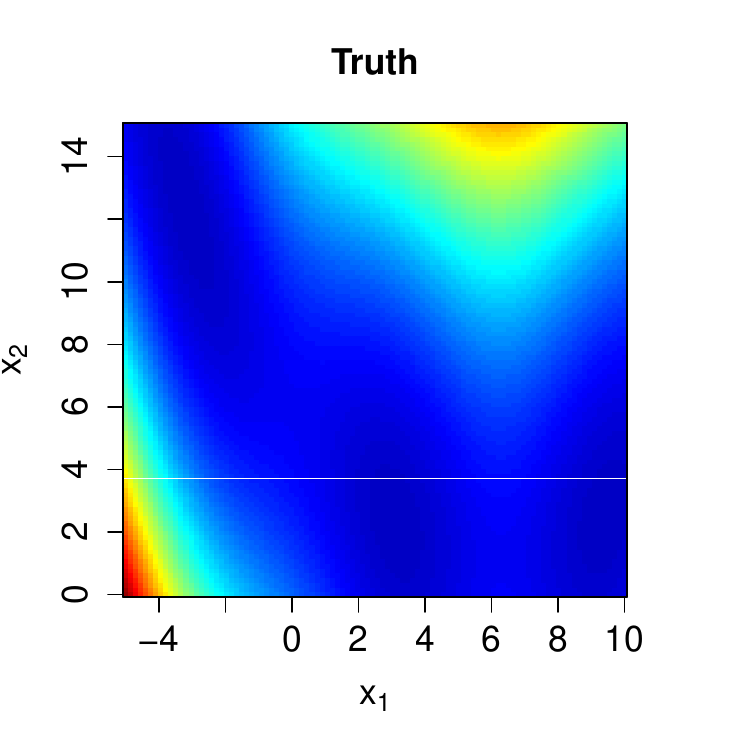}
 \hspace{-.25in}        \includegraphics[width=.33\textwidth]{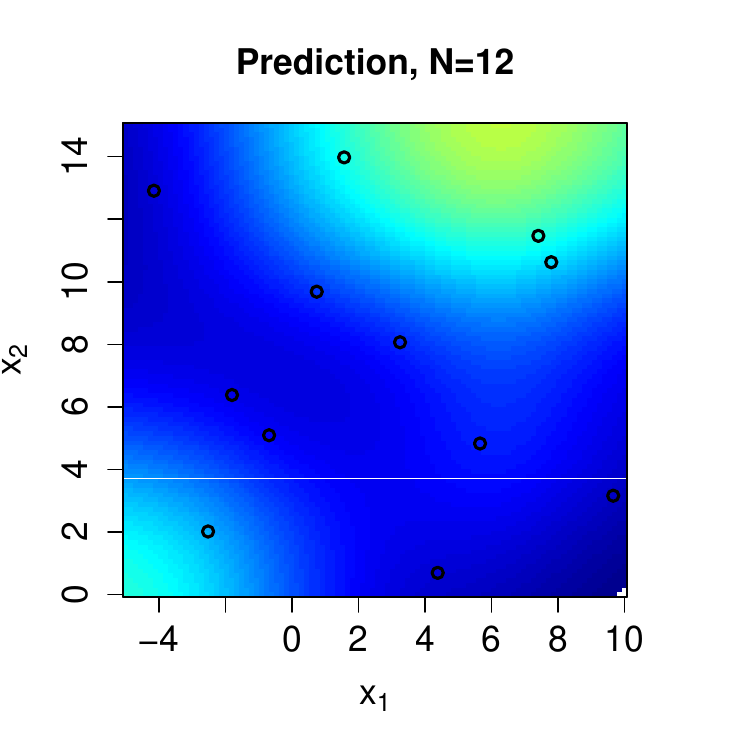}  
\hspace{-.25in}        
 \includegraphics[width=.33\textwidth]{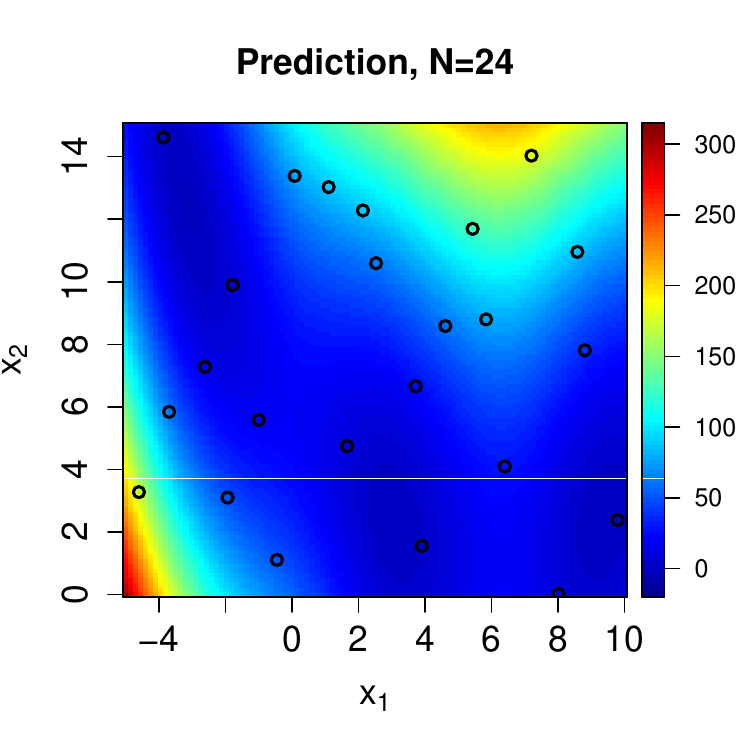}
        
        \vspace{-.2in}
  \end{tabular}
   \caption{Predictions by the GP emulator of a function on 2D inputs with $N=12$ and $N=24$ observations (black circles) are shown in the left and right panels, respectively. }
\label{fig:branin}
\end{figure}

Constructing a GP emulator to approximate computer simulation typically starts with a ``space-filling" design, such as the Latin hypercube sampling (LHS), to fill the input space. Numerical solutions of computer models were then obtained  at these design points, and the set $\{(\mathbf x_i,y_i)\}^{N}_{i=1}$ is used for training a GP emulator.  For any observed input $\mathbf x^*$, the predictive mean  in (\ref{equ:predictiongp}) is often used for predictions, and the uncertainty of observations can be obtained through the predictive distribution. 
in Figure \ref{fig:branin}, we plot the predictive mean of a GP emulator to approximate the Branin function  \cite{simulationlib} with N training inputs sampled from LHS,  using the default setting of the {\tt RobustGaSP} package \cite{gu2018robustgasp}.
 When the number of observations increases from $N=12$ (middle panel) to $N=24$ (right panel), the predictive error becomes smaller.

The computational complexity of GP models increases at the order of $\mathcal O(N^3)$, which prohibits applications on emulating complex computer simulations, when  a relatively large number of simulation runs are required. In the next section, we will introduce the state space representation of GP with Mat{\'e}rn covariance and one-dimensional input, where the computational order scales as $\mathcal O(N)$ without approximation. This method can be applied to problems with high dimensional input space discussed in Section \ref{sec:particle_interactions}.

\section{Marginalization in Kalman filter} 
\label{eq:marginalization_KF}
\subsection{State space representation of GP with the Mat{\'e}rn kernel}
\label{subsec:state_space}
Suppose we model the observations by Equation (\ref{equ:obs_noise}) where  the latent process $z(.)$ is assumed to follow a GP on 1D input. For simplicity, here we assume a zero mean parameter ($\mu=0$), and a mean function can be easily included in the analysis. It has been realized that a GP defined in 1D input space using a Mat{\'e}rn covariance  with a half-integer roughness parameter input can be written as stochastic differential equations (SDEs) \cite{whittle1963stochastic,hartikainen2010kalman}, which can reduce the  operations of computing the likelihood and making predictions from $\mathcal O(N^3)$  to $\mathcal O(N)$ operations, with the use of Kalman filter. Here we first review SDE representation and then we discuss marginalization  of latent variables in the Kalman filter algorithm for scalable computation. 
 
When the roughness parameter is $\nu=5/2$, for instance, the Mat{\'e}rn kernel has the expression below
 \begin{equation}
 K(d)= \left(1+\frac{\sqrt{5}d}{\gamma}+\frac{5d^2}{3\gamma^2}\right)\exp\left(-\frac{\sqrt{5}d}{\gamma}\right), \,
\label{equ:matern_5_2}
\end{equation}
where $d=|x_a-x_b|$ is the distance between any $x_a,x_b \in \mathbb R$ and $\gamma$ is a range parameter typically estimated by data. 
The output and two derivatives of the GP with the Mat{\'e}rn kernel in (\ref{equ:matern_5_2}) can be written as the SDE below \cite{hartikainen2010kalman},  
\begin{align}
  \frac{d\bm {\theta}(x)}{dx}=\mathbf J\bm {\theta} (x)+\sqrt{c}\mathbf L \varepsilon(x),  
  \label{equ:SDE_matern_5_2}
  \end{align}
or in the matrix form, 
\begin{equation*}
 \frac{d}{dx} \begin{pmatrix}
 z(x) \\ 
 z^{(1)}(x)  \\ 
 z^{(2)}(x) 
\end{pmatrix}= 
\begingroup 
\setlength\arraycolsep{2pt}
\begin{pmatrix}
 0&1  &0 \\ 
 0&0  &1 \\ 
 -\lambda^3& -3\lambda^2  &-3\lambda
\end{pmatrix} 
\endgroup
\begin{pmatrix}
 z(x) \\ 
 z^{(1)}(x)  \\ 
 z^{(2)}(x) 
\end{pmatrix}+\sqrt{c}\begin{pmatrix}
0 \\ 
0  \\ 
1
\end{pmatrix}  \varepsilon(x),
\end{equation*}
where $\varepsilon(x)$ is a  standard Gaussian white noise process,  $c=\frac{16}{3}\sigma^2\lambda^5$, $\lambda=\frac{\sqrt{2\nu}}{\gamma}=\frac{\sqrt{5}}{\gamma}$, and  $z^{(l)}(\cdot)$ is the $l$th derivative of the process $z(\cdot)$. Denote $\mathbf F=(1,0,0)$. Assume the 1D input is ordered, i.e. $x_1< x_2<...< x_N$.   The solution of  SDE in (\ref{equ:SDE_matern_5_2}) can be expressed as a continuous-time dynamic linear model \cite{West1997},  
\begin{align}
\label{equ:ctdlm}
\begin{split}
y(x_i)&= \mathbf F\bm \theta(x_i) + \epsilon(x_i), \\
\bm \theta(x_i)&=\mathbf G(x_i) \bm \theta(x_{i-1}) +\mathbf w(x_i), \,    
\end{split}
\end{align}
where $\mathbf w(x_i) \sim \mathcal{MN}(0, \mathbf W(x_i))$ for $i=2,...,N$,    $\bm \theta(x_1) \sim \mathcal{MN}(\mathbf 0, \mathbf W(x_1))$ and Gaussian noise follows $\epsilon(x_i)\sim \mathcal{N}(0,\sigma^2_0)$. Here
 $\mathbf G(x_i)=e^{\mathbf J(x_{i}-x_{i-1})}$ and $\mathbf W(x_i) =\int^{x_{i}-x_{i-1}}_0 e^{\mathbf J t} \mathbf L c \mathbf L^T e^{\mathbf J^T t} dt$ from $i=2,...,N$, and stationary distribution $\bm \theta(x_{i}) \sim \mathcal{MN}(0, \mathbf W(x_1) ) $, with $\mathbf W(x_1)= \int^{\infty}_{0} e^{\mathbf Jt}\mathbf Lc\mathbf L^Te^{\mathbf J^T t}dt.$ 
Both $\mathbf G(x_i) $ and $\mathbf W(x_i)$ have closed-form expressions  given in the Appendix \ref{sec:close_formed_state_space}. The joint distribution of the states follows  $\left(\bm \theta^T(x_1),...,\bm \theta^T(x_N) \right)^T\sim \mathcal{MN}(\mathbf 0, \bm \Lambda^{-1} )$, where the  inverse covariance $\bm \Lambda$  is a block tri-diagonal matrix discussed in Appendix \ref{sec:close_formed_state_space}.

\subsection{Kalman filter as a scalable marginalization technique}
\label{subsec:KF}

For dynamic linear models in (\ref{equ:ctdlm}), Kalman filter  and Rauch–Tung–Striebel (RTS) smoother can be  used as an exact and scalable approach to compute the likelihood, and predictive distributions.  The Kalman filter and RTS smoother are sometimes called the forward filtering and backward smoothing/sampling algorithm, widely used in dynamic linear models of time series. We refer the readers to  \cite{West1997,petris2009dynamic} for  discussion of dynamic linear models. 

Write $\mathbf G(x_i)=\mathbf G_{i} $, $\mathbf W(x_i)=\mathbf W_i $, $\bm \theta(x_i)=\bm \theta_i$ and $y(x_i)=y_i$ for $i=1,...,N$. In  Lemma \ref{lemma:KF}, we summarize  Kalman filter and RTS smoother for the dynamic linear model in (\ref{equ:ctdlm}). Compared with $\mathcal O(N^3)$ computational operations and $\mathcal O(N^2)$ storage cost from GPs, the outcomes of Kalman filter and RTS smoother can be used  for computing the likelihood and predictive distribution with $\mathcal O(N)$ operations and $\mathcal O(N)$ storage cost, summarized in Lemma \ref{lemma:KF}. All the distributions in Lemma \ref{lemma:KF} and Lemma \ref{lemma:lik_pred_KF} are  conditional distributions given parameters  $(\gamma, \sigma^2, \sigma^2_0)$, which are omitted for simplicity.

\begin{lemma}[Kalman Filter and RTS Smoother \cite{kalman1960new,rauch1965maximum}]
\label{lemma:KF}
~~~

  \noindent 1. ({\bf Kalman Filter}.) Let $\bm \theta_{i-1}| \mathbf y_{1:i-1} \sim  \mathcal{MN}( \mathbf m_{i-1}, \mathbf C_{i-1})$. For $i=2,...,N$, iteratively we have, 
\begin{itemize}
\item[(i)] the one-step-ahead predictive distribution of $\bm \theta_i$ given  $\mathbf y_{1:i-1}$,  
 \vspace{-.05in}
 \begin{equation}
 \bm \theta_i| \mathbf y_{1:i-1} \sim  \mathcal{MN}(\mathbf b_i, \mathbf B_i), 
  \vspace{-.05in}
 \label{equ:KF1}
 \end{equation}
with $\mathbf b_i= \mathbf G_i \mathbf m_{i-1} $ and $\mathbf B_i= \mathbf G_i \mathbf C_{i-1} \mathbf G^T_i+\mathbf W_i$,  
\item[(ii)] the one-step-ahead predictive distribution of $Y_i$ given $\mathbf y_{1:i-1}$,
 \vspace{-.05in}
 \begin{equation}
Y_i| \mathbf y_{1:i-1} \sim  \mathcal{N}(f_i, Q_i),
 \vspace{-.05in}
 \label{equ:KF2}
\end{equation}
with   $f_i= \mathbf F \mathbf b_{i}, $ and $Q_i= \mathbf F \mathbf B_i \mathbf F^T+ \sigma^2_0$, 
\item[(iii)] the filtering distribution of $\bm \theta_i$ given $\mathbf y_{1:i}$, 
 \vspace{-.05in}
 \begin{equation}
\bm \theta_i| \mathbf y_{1:i} \sim  \mathcal{MN}(\mathbf m_i, \mathbf C_i),  
 \vspace{-.05in}
 \label{equ:KF3}
\end{equation}
with  $\mathbf m_i= \mathbf b_i + \mathbf B_i \mathbf F^T Q^{-1}_i  (y_i -f_i)$ and  $\mathbf C_i= \mathbf B_i -  \mathbf B_i \mathbf F^T   Q^{-1}_i \mathbf F \mathbf B_i$. 
\end{itemize}

\noindent 2. ({\bf RTS Smoother}.)  Denote $\bm \theta_{i+1}| \mathbf y_{1:n} \sim  \mathcal{N}(s_{i+1}, S_{i+1})$, then recursively for $i=N-1,...,1$, 
 \vspace{-.05in}
  \begin{equation}
 \bm \theta_i| \mathbf y_{1:N} \sim  \mathcal{MN}( \mathbf s_i, \mathbf S_i), 
 \vspace{-.05in}
  \label{equ:KF4}
 \end{equation}
 {where  $\mathbf s_i=\mathbf m_i +  \mathbf C_i \mathbf G^T_{i+1} \mathbf B^{-1}_{i+1}(\mathbf s_{i+1} -\mathbf b_{i+1})$ and  $\mathbf S_i = \mathbf C_i- \mathbf C_i \mathbf G^T_{i+1} \mathbf B^{-1}_{i+1} ( \mathbf B_{i+1} -\mathbf S_{i+1})\mathbf B^{-1}_{i+1}\mathbf G_{i+1} \mathbf C_i$}.

\end{lemma}

\begin{lemma}[Likelihood and predictive distribution]
\label{lemma:lik_pred_KF}

~~~

\noindent 1. ({\bf Likelihood}.) The likelihood follows 
\begin{equation*}
p(\mathbf y_{1:N} \mid \sigma^2, \sigma_0^2, \gamma)=\left\{\prod^N_{i=1}(2\pi Q_i)^{-\frac{1}{2}}\right\} \exp\left\{-\sum^{N}_{i=1}\frac{ (y_i  -f_i)^2 }{2Q_i} \right\},
\label{equ:KF_lik}
\end{equation*}
where $f_i$ and $  Q_i$ are given in Kalman filter.   The likelihood can be used to obtain the MLE of the parameters $(\sigma^2, \sigma_0^2, \gamma)$. 

\noindent 2. ({\bf Predictive distribution}.) 
\begin{itemize}
\item[(i)] By the last step of Kalman filter, one has $\bm \theta_N| \mathbf y_{1:N}$ and recursively by the RTS smoother, the predictive distribution of $\bm \theta_i$ for $i=N-1,...,1$ follows
 \vspace{-.05in}
  \begin{equation}
 \bm \theta_i| \mathbf y_{1:N} \sim  \mathcal{MN}( \mathbf s_i, \mathbf S_i).
 \vspace{-.05in}
  \label{equ:KF_pred}
 \end{equation}
 \item[(ii)] For any $x^*$  (W.l.o.g. let $x_i< x^* < x_{i+1}$) 
\[\bm \theta(x^*) \mid  \mathbf y_{1:N} \sim \mathcal{MN}\left(\hat {\bm \theta}(x^*), \hat {\bm \Sigma}(x^*)  \right) \]
 where 
 \begin{align*}
 \hat {\bm \theta}(x^*)&=\mathbf G^*_i \mathbf s_{i}+ \mathbf W^*_i (\mathbf G^*_{i+1})^T (\tilde{\mathbf W}^*_{i+1})^{-1}(\mathbf s_{i+1}- \mathbf G^*_{i+1}\mathbf G^*_i \mathbf s_{i})  \\ \hat {\bm \Sigma}(x^*)&=( (\mathbf W^*_{i})^{-1}+(\mathbf G^*_{i+1})^T (\mathbf W^*_{i+1})^{-1} \mathbf G^*_{i+1} )^{-1} 
 \end{align*} 
 with terms denoted with `*'  given in the Appendix \ref{sec:close_formed_state_space}.  

 \end{itemize}
\end{lemma}

Although we introduce the Mat{\'e}rn kernel with $\nu=5/2$ as an example, the likelihood and predictive distribution of GPs with the Mat{\'e}rn kernel of a small half-integer roughness parameter  can be computed efficiently, for both equally spaced and not equally spaced 1D inputs. For the Mat{\'e}rn kernel with a very large roughness parameter, the dimension of the latent states becomes large, which makes efficient computation prohibitive. In practice, the Mat{\'e}rn kernel with a relatively large roughness parameter (e.g. with $\nu=5/2$) is found to be accurate for estimating a smooth latent function in computer experiments \cite{Gu2018robustness,anderson2019magma}. Because of this reason, the Mat{\'e}rn kernel with $\nu=5/2$ is the default choice of the kernel function in some packages of GP emulators   \cite{roustant2012dicekriging,gu2018robustgasp}.

For a model containing latent variables, one may proceed with two usual approaches:
\begin{itemize}
\item[(i)]   sampling the latent variables $\bm \theta({x_i})$ from the posterior distribution by the MCMC algorithm,
\item[(ii)]   optimizing  the latent variables $\bm \theta({x_i})$ to minimize a loss function. 
\end{itemize}
For approach (i),  the MCMC algorithm is usually much slower than the Kalman filter, as the number  of the latent states 
is high, requiring a large number of posterior samples \cite{10.1214/21-BA1295}. On the other hand, the prior correlation between states may not be taken into account directly in approach (ii), making the estimation less efficient than the Kalman filter, if data contain correlation across latent states. In comparison, the latent states in the dynamic linear model in (\ref{equ:ctdlm})  are iteratively marginalized out in Kalman filter, and the closed-form expression is derived in each step, which only takes $\mathcal O(N)$ operations and storage cost, with $N$ being the number of observations. 

In practice, when a sensible probability model or a prior of latent variables is considered,  the principle is to integrate out the latent variables when  making predictions. Posterior samples and optimization algorithms, on the other hand, can be very useful for approximating the marginal likelihood when closed-form expressions are not available. As an example, we will introduce applications that integrate the sparse covariance structure along with conjugate gradient optimization into estimating particle interaction kernels, and forecasting particle trajectories in Section \ref{sec:particle_interactions}, which integrates both marginalization and optimization to tackle a computationally challenging scenario.

\begin{figure}[t] 
\centering
  \begin{tabular}{cc}
 \hspace{-.25in}  
    \includegraphics[width=.5\textwidth]{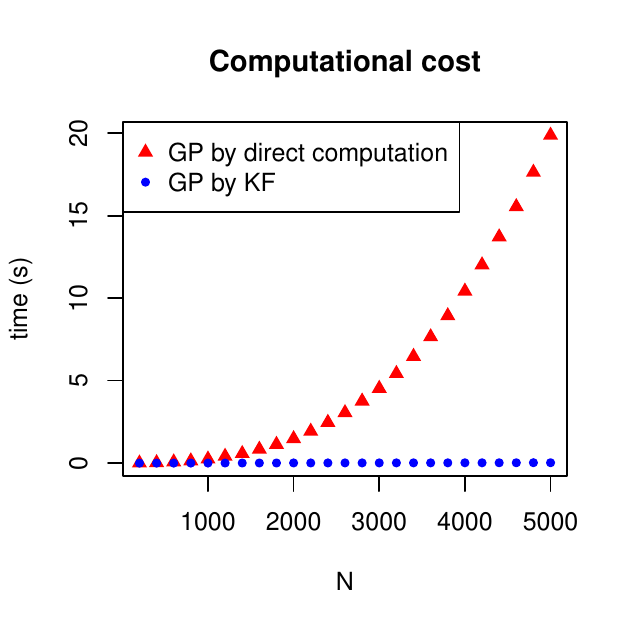}
 \hspace{-.3in}      
 \includegraphics[width=.5\textwidth]{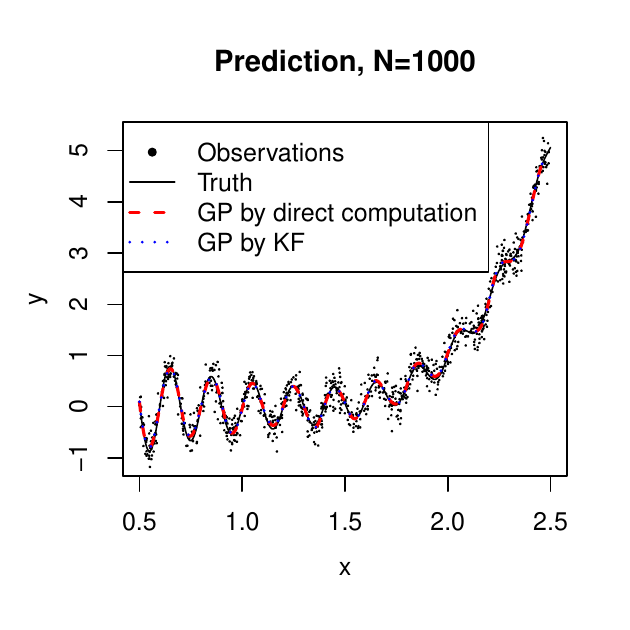}

 \vspace{-.2in}
  \end{tabular}
   \caption{Comparison between the direct computation of GP, and that by the Kalman filter (KF) and  RTS smoother, both having the Mat{\'e}rn kernel in (\ref{equ:matern_5_2}). The left panel shows the computational cost of computing the predictive mean over different number of observations. When $N=5000$, direct computation takes around 20 seconds, whereas the computation by KF and RTS smoother takes 0.029 seconds. The predictive mean computed in both ways is plotted in the right panel when $N=1,000$, where the root of mean squared difference between two approaches is $5.98\times 10^{-12}$.  }
\label{fig:Comparison_KF_GP}
\end{figure}

In Figure \ref{fig:Comparison_KF_GP}, we compare the cost for computing the predictive mean for a nonlinear function with 1D inputs \cite{gramacy2012cases}. The input is uniformly sampled from  $[0.5,0.25]$, and an independent Gaussian white noise with a standard deviation of $0.1$ is added in simulating the observations. We compare two ways of computing the predictive mean. The first approach implements direct computation of the predictive mean by Equation (\ref{equ:gppredmean}). The second approach is computed by the likelihood function and predictive distribution from Lemma \ref{lemma:lik_pred_KF} based on the  Kalman filter and  RTS smoother. The range and nugget parameters are fixed to be $0.5$ and $10^{-4}$ for demonstration purposes, respectively. The computational time of this simulated experiment is shown in the left panel in Figure \ref{fig:Comparison_KF_GP}. The approach based on Kalman filter and RTS smoother is much faster, as  computing the likelihood and making predictions by Kalman filter and  RTS smoother only require $\mathcal O(N)$    operations, whereas the direct computation cost $\mathcal O(N^3)$ operations. The right panel gives the predictive mean, latent truth, and observations, when  $N=1000$. The difference between the two approaches is very small, as both methods are exact.

\subsection{Marginalization of correlated matrix observations with multi-dimensional inputs}
\label{subsec:marginalization_multidim}
The Kalman filter is widely applied in 
signal processing, system control, and modeling time series. Here we introduce a few recent studies that apply Kalman filter to GP models with Mat{\'e}rn covariance to model  spatial, spatio-temporal, and functional observations. 

Let $\mathbf y(\mathbf x)= (y_1(\mathbf x),...,y_{n_1}(\mathbf x))^T$ be an $n_1$-dimensional real-valued output vector at a $p$-dimensional input vector $\mathbf x$. For simplicity, assume the mean is zero. Consider the latent factor model:  
	\begin{equation}
	\mathbf y(\mathbf x)= \mathbf A \mathbf z(\mathbf x) + \bm \epsilon(\mathbf x),
	\label{equ:model_latent_factor}
	\end{equation}
where  $\mathbf A=[\mathbf a_1,...,\mathbf a_d]$ is an $n_1\times d$ factor loading matrix and  $\mathbf z(\mathbf x)=(z_1(\mathbf x),...,z_d(\mathbf x))^T$ is a $d$-dimensional factor processes, with $d \leq n_1$. The noise process follows $\bm \epsilon(\mathbf x) \sim \mathcal N(\mathbf 0,\, \sigma^2_0 \mathbf I_{n_1})$. Each factor is modeled  by a zero-mean Gaussian process (GP), meaning that $\mathbf Z_l=(z_l(\mathbf x_1),...,z_l(\mathbf x_{n_2}))$ follows a multivariate normal distribution  $\mathbf Z^T_l \sim \mathcal{MN}(\mathbf 0, \bm \Sigma_l)$,	where the $(i,\, j)$ entry of $\bm \Sigma_l$ is parameterized by a covariance function $\sigma^2_l K_l(\mathbf x_i, \mathbf x_j)$ for $l=1,...,d$. The model (\ref{equ:model_latent_factor}) is often known as the {semiparametric latent factor model} in the machine learning community \cite{seeger2005semiparametric}, and it belongs to a class of {linear models of coregionalization}  \cite{banerjee2014hierarchical}. It has a wide range of applications in modeling multivariate spatially correlated data and functional observations from computer experiments \cite{gelfand2005spatial,higdon2008computer,paulo2012calibration}.

We have the following two assumptions for model (\ref{equ:model_latent_factor}). 

\begin{assumption}
The prior of latent processes $\mathbf{z}_i(.)$ and $\mathbf{z}_j(.)$ are independent, for any $i\neq j$.
\label{assump:1}
\end{assumption}

\begin{assumption}
The factor loadings are orthogonal, i.e. $\mathbf A^T\mathbf A=\mathbf I_d$. 
\label{assump:2}
\end{assumption}
The first assumption is typically assumed for modeling multivariate spatially correlated data or computer experiments \cite{banerjee2014hierarchical,higdon2008computer}. Secondly, note that the model in (\ref{equ:model_latent_factor}) is unchanged if we replace    $(\mathbf A, \mathbf z(\mathbf x))$ by $(\mathbf A \mathbf E, \mathbf E^{-1} \mathbf z(\mathbf x))$  for an invertible  matrix $\mathbf E$, meaning that the linear subspace of $\mathbf A$ can be identified if no further constraint is imposed. Furthermore, as the variance of each latent process $\sigma^2_i$ is estimated by the data, imposing the unity constraint on each column of $\mathbf A$ can reduce identifiability issues. The second assumption was also assumed in other recent studies  \cite{lam2011estimation,lam2012factor}.

Given Assumption \ref{assump:1} and Assumption \ref{assump:2}, we review recent results that alleviates the computational cost. Let us first assume the observations are an $N=n_1\times n_2$ matrix $\mathbf Y=[\mathbf y(\mathbf x_1),...,\mathbf y(\mathbf x_{n_2})]$.

\begin{lemma}[Posterior independence and orthogonal projection \cite{10.1214/21-BA1295}]
For model (\ref{equ:model_latent_factor}) with Assumption \ref{assump:1} and Assumption \ref{assump:2}, we have two properties below. 

\noindent 1. ({\bf Posterior Independence}.) For any   $l\neq m$
    \[\mbox{Cov}[\mathbf{Z}_l^T, \mathbf{Z}_m^T | \mathbf{Y}] = \mathbf{0},\]
    and for each $l=1,...,d$, $$\mathbf{Z}_l^T|\mathbf{Y}\sim  \mathcal{MN}(\bm{\mu}_{Z_l}, \bm{\Sigma}_{z_l}),$$
    where $\bm{\mu}_{z_l} = \bm{\Sigma}_l \bm{\tilde{\Sigma}}^{-1}_l \mathbf{\tilde{y}}_l$, $\mathbf{\tilde{y}}_l = \mathbf{Y}^T \mathbf{a}_l$ and $\bm{\Sigma}_{Z_l} = \bm{\Sigma}_l - \bm{\Sigma}_l \bm{\tilde{\Sigma}}_l^{-1} \bm{\Sigma}_l$ with $\bm{\tilde{\Sigma}}_l = \bm{\Sigma}_l + \sigma_0^2 \mathbf{I}_{n_2}$.
\label{lemma:post_projection}   

\noindent 2. (\textbf{Orthogonal projection}.) After integrating $\mathbf \mathbf{z}(\cdot)$,  the marginal likelihood  is a product of multivariate normal densities at projected observations:
	\begin{equation}
	p(\mathbf Y)= \prod^{d}_{l=1} \mathcal{PN}(\tilde {\mathbf y}_l; \mathbf 0, \bm {\tilde \Sigma}_l ) \prod^{n_1}_{l=d+1} \mathcal{PN}(\tilde {\mathbf y}_{c,l}; \mathbf 0, \sigma^2_0 \mathbf I_{n_2}),  
	\label{equ:marginal_lik}
	\end{equation}
	where $\tilde {\mathbf y}_{c,l} = \mathbf Y^T  \mathbf a_{c,l} $ with $\mathbf a_{c,l}$ being the $l$th column of  $\mathbf A_c$, the orthogonal component of $\mathbf A$, and $\mathcal{PN}$ denotes the density for a multivariate normal distribution.

\end{lemma}

The  properties in Lemma \ref{equ:model_latent_factor} lead to computationally scalable expressions of the maximum marginal likelihood estimator (MMLE) of factor loadings.   

	\begin{theorem}[Generalized probabilistic principal component analysis \cite{gu2018generalized}]
	Assume $\mathbf A^T \mathbf A=\mathbf I_d$, after marginalizing out $\mathbf Z^T_l \sim \mathcal{MN}(\mathbf 0, \bm \Sigma_l)$ for $l=1,2,...,d$, we have the results below. 
	\begin{itemize}
	\item If $\bm \Sigma_1=...=\bm \Sigma_d=\bm \Sigma$, the marginal  likelihood  is maximized when  
	\begin{equation}
	 \hat {\mathbf A}=\mathbf U \mathbf S, 		
	 	\label{equ:A_est_shared_cov}
	\end{equation}
	 where $\mathbf U$ is an $n_1 \times d$ matrix of the first $d$ principal eigenvectors of $\mathbf G={\mathbf Y (\sigma^2_0 \bm \Sigma^{-1}+  \mathbf I_{n_2} )^{-1}  \mathbf Y^T}$ 
	and $\mathbf S$ is a $d \times d$ orthogonal rotation matrix;
	\item If the covariances of the factor processes are different, denoting $\mathbf G_l= { \mathbf Y  (\sigma^2_0 \bm \Sigma^{-1}_l+\mathbf I_{n_2} )^{-1}\mathbf Y^T}$, the MMLE of factor loadings is 
	     	\begin{equation}
\mathbf {\hat A}= \argmax_{\mathbf A} \sum^d_{l=1}  {\mathbf a^T_l \mathbf G_l \mathbf a_l},  \quad \text{s.t.} \quad \mathbf A^T \mathbf A=\mathbf I_d. 
	\label{equ:A_est_diff_cov}
	\end{equation}
	\end{itemize}
		\label{thm:est_A}
		\end{theorem}
The estimator  $\mathbf A$ in Theorem \ref{thm:est_A}  is called the \textit{generalized probabilistic principal component analysis} (GPPCA). The optimization algorithm that preserves the orthogonal constraints in (\ref{equ:A_est_diff_cov}) is available in \cite{wen2013feasible}.

In \cite{tipping1999probabilistic},  the latent factor is assumed to follow independent standard normal distributions, and the authors derived the MMLE of the factor loading matrix $\mathbf A$, which was termed the probabilistic principal component analysis (PPCA). The GPPCA extends the PPCA to correlated  latent factors modeled by GPs, which incorporates the prior correlation information between outputs as a function of inputs, and the latent factor processes were marginalized out when estimating the factor loading matrix and other parameters. When the input is 1D and the Mat{\'e}rn covariance is used for modeling latent factors, the computational order of GPPCA is $\mathcal O(Nd)$, which is the same as the PCA. For correlated data, such as spatio-temporal observations and  multivariate functional data, GPPCA provides a flexible and scalable approach to estimate factor loading by marginalizing out the latent factors \cite{gu2018generalized}.

Spatial and spatio-temporal models with a separable covariance can be written as a special case of the model in (\ref{equ:model_latent_factor}). For instance, suppose $d=n_1$ and the $n_1\times n_2$ latent factor matrix follows $\mathbf Z\sim \mathcal{MN}(0,\, \sigma^2 \mathbf R_1 \otimes \mathbf R_2 ),$ where $\mathbf R_1$ and $\mathbf R_2$ are $n_1\times n_1$ and $n_2\times n_2$ subcovariances, respectively. Denote the eigen decomposition  $\mathbf R_1=\mathbf V_1 \bm \Lambda_1 \mathbf V_1^T$ with $ \bm \Lambda_1$ being a diagonal matrix with the eigenvalues $\lambda_i$, for $i=1,...,n_1$. Then this separable model can be written as the model in (\ref{equ:model_latent_factor}), with $\mathbf A=\mathbf V_1$, $\bm \Sigma_l=\sigma^2 \lambda_{l} \mathbf R_2$. The connection suggests that the latent factor loading matrix can be specified as the eigenvector matrix of a covariance matrix, parameterized by a kernel function. This approach is studied in \cite{10.1214/21-BA1295} for modeling incomplete lattice with irregular missing patterns, and the Kalman filter is integrated for accelerating computation on massive spatial and spatio-temporal observations.

\section{Scalable marginalization for learning particle interaction kernels from trajectory data} 
\label{sec:particle_interactions}
Collective motions with particle interactions are very common in both microscopic and macroscopic systems \cite{marchetti2013hydrodynamics,motsch2014heterophilious}. Learning interaction kernels between particles is important for two purposes. First, physical laws are less understood for many complex systems, such as cell migration  or non-equilibrium thermodynamic processes. Estimating the interaction kernels between particles from fields or experimental data  is essential for learning these processes. Second,  simulation of particle interactions, such as \textit{ab initio} molecular dynamics   simulation,  can be very computationally expensive. Statistical learning approaches can be used for reducing the computational cost of simulations.  

For demonstration purposes, we consider a simple first-order system. In \cite{lu2019nonparametric}, for a system with $n$ interacting particles, the velocity of the $i$th particle at time $t$, $\mathbf v_i(t)=d  \mathbf x_i(t)/d t$, is modeled by positions between all particles, 
\begin{equation}
 \mathbf v_i(t)=\sum^n_{j=1} \phi(||\mathbf x_j(t)-\mathbf x_i(t) ||) \mathbf u_{i,j}(t),  
  \label{equ:interaction}
\end{equation}
where $\phi(\cdot)$ is a latent interaction kernel function between particle $i$ and all other particles, with  $|| \cdot ||$ being the Euclidean distance, and $\mathbf u_{i,j}(t)=\mathbf x_j(t)-\mathbf x_i(t)$ is a vector of differences between positions of particles $i$ and $j$, for $i,j=1,...,n$. Here $\phi(\cdot)$ is a two-body interaction function. In molecular
dynamics simulation, for instance, the Lennard-Jones
potential can be related to molecular forces and the acceleration
of molecules in a second-order system. The statistical learning approach can be extended to a second-order system that involves   acceleration and  external force terms.  The first-order system as  (\ref{equ:interaction}) can be considered as an approximation of the second-order system.  Furthermore, the interaction between particles is global, as any particle is affected by all other particles. Learning global interactions  is more computationally challenging than local interactions \cite{sanchez2020learning}, and approximating the global interaction by the local interaction is of interest in future studies.

One important goal is to efficiently estimate the unobservable interaction functions from the particle trajectory data, without specifying a parametric form. This goal is key for estimating the behaviors of dynamic systems in experiments and in observational studies, as the physics law in a new system may be unknown. 
In \cite{feng2021data}, $\phi(\cdot)$ is modeled by a zero-mean Gaussian process with a Mat{\'e}rn covariance: 
 \begin{equation}
 \phi(\cdot) \sim \mathcal{GP}(0, \sigma^2 K(\cdot,\cdot) ). 
 \label{equ:phi}
 \end{equation}
Computing estimation of interactions of large-scale systems or more simulation runs, however, is prohibitive, as the direct inversion of the covariance matrix of  observations of velocities requires $\mathcal O((nMDL)^3)$ operations,  where   $M$ is the number of  simulations or experiments,  $n$ is the number of particles, $D$ is the dimension of each particle,  $L$ denotes the number of time points for each trial. Furthermore, constructing such covariance   contains computing an $n^2LM \times n^2LM$ matrix of $\phi$ for a $D$-dimensional input space, which takes $\mathcal O(n^4L^2M^2D)$ operations.  Thus,  directly estimating interaction kernel with a GP model in (\ref{equ:phi})  is only applicable to systems with a  small number of observations  \cite{lu2019nonparametric,feng2021data}.
 
This work makes contributions from two different aspects for estimating dynamic systems of interacting particles. We first show the covariance of velocity observations  can be obtained by  operations on a few sparse matrices,  after marginalizing out the latent interaction function. The sparsity of the inverse covariance of the latent interaction kernel allows us to modify the Kalman filter  to efficiently compute the matrix product in this problem, and then apply a conjugate gradient (CG) algorithm \cite{hestenes1952methods,hackbusch1994iterative,saad2003iterative} to solve this system iteratively.  
 The computational complexity of computing the predictive mean and variance of a test point is at the order of $\mathcal O(TnN)+\mathcal O(nN\log(nN))$, for a system of $n$ particles, $N=nMDL$ observations, and $T $ is the number of iterations required in the CG algorithm. We found that typically around a few hundred CG iterations can achieve high predictive accuracy for a moderately large number of observations. The algorithm leads substantial reduction of computational cost, compared to direct computation. 
 
 Second, we study the effect of experimental designs on estimating the interaction kernel function. In previous studies, it is unclear how initial positions, time steps of trajectory and the number of particles affect the accuracy in estimating  interaction kernels. 
Compared to other conventional problems in computer model emulation, where a ``space-filling" design is often used, here we cannot directly observe the realizations of the latent function. Instead, the output velocity is a weighted average of the interaction kernel functions between particles. Besides, we cannot directly control distances between the particles moving away from the initial positions, in both simulation and experimental studies. When the distance between two particles $i$ and $j$ is small,  the contribution  $\phi(||\mathbf x_i(t)-\mathbf x_j(t) ||) \mathbf u_{i,j}(t)$ can be small, if the  repulsive force by $\phi(\cdot)$ does not increase as fast as the distance decreases. Thus we found that the estimation of interaction kernel function can be less accurate in the input domain of small distances. This problem can be alleviated by placing initial positions of more particles close to each other, providing more data with small distance pairs that improve the accuracy in estimation.

\subsection{Scalable computation based on sparse representation of inverse covariance}
For illustration purposes, let us first consider a simple scenario where we have $M=1$ simulation and $L=1$ time point of a system of $n$ interacting particles at a $D$ dimensional space.  Since we only have 1 time point, we omit the notation $t$ when there is no confusion. The algorithm for the general scenario with $L>1$ and $M>1$ is discussed in Appendix \ref{subsec:algorithm_particles}.  In practice, the experimental observations of velocity from  multiple particle tracking algorithms or  particle image velocimetry  typically contain noises \cite{adrian2011particle}. Even for simulation data, the numerical error could be  non-negligible for large and complex systems. In these scenarios, the observed velocity $\mathbf {\tilde v}_i=(v_{i,1},...,v_{i,D})^T$  is a noisy realization: 
$\mathbf {\tilde v}_i=\mathbf v_i+\bm \epsilon_i $, where  $\bm \epsilon_i \sim \mathcal{MN}(0,\sigma^2_0 \mathbf I_D)$ denotes a vector of Gaussian noise with  variance $\sigma^2_0$. 

Assume the $nD$ observations of velocity are $\mathbf {\tilde v}=(\tilde v_{1,1},...,\tilde v_{n,1},\tilde v_{1,2},...,\tilde v_{n,2},...,\tilde v_{n-1,D},\tilde v_{n,D})^T$. 
After integrating out the latent function modeled in Equation (\ref{equ:phi}), the marginal distribution of observations follows 
\begin{align}
(\mathbf {\tilde v} \mid  \mathbf R_{\phi}, \sigma^2, \sigma^2_0) \sim \mathcal{MN}\left(\mathbf 0,  \sigma^2 \mathbf U \mathbf R_{\phi} \mathbf U^T +\sigma^2_0 \mathbf I_{nD}  \right),
\label{equ:v}
\end{align}
where $\mathbf U$ is an $nD \times n^2$ block diagonal matrix, with the $i$th $D \times n$ block in the diagonals being $(\mathbf u_{i,1},...,\mathbf u_{i,n})$, and $\mathbf R_{\phi}$ is an $n^2\times n^2$ matrix, where the $(i',j')$ term in the $(i,j)$th $n\times n$ block is $K(d_{i,i'},d_{j,j'} )$ with $d_{i,i'}=||\mathbf x_i-\mathbf x_i' ||$ and $d_{j,j'}=||\mathbf x_j-\mathbf x_j' ||$ for $i,i',j,j'=1,...,n$. Direct computation of the likelihood involves computing the inverse of an $nD\times nD$ covariance matrix and constructing an $n^2\times n^2$ matrix $\mathbf R_{\phi}$, which costs $\mathcal O((nD)^3)+\mathcal O(n^4D)$ operations. This is  computationally expensive even for small systems.  

Here we use an exponential kernel function, $K(d)=\exp(-d/\gamma)$ with range parameter $\gamma$, of modeling any nonnegative distance input $d$ for illustration, where this method can be extended to include Mat{\'e}rn kernels with half-integer roughness parameters. Denote distance pairs $d_{i,j}=||\mathbf x_i-\mathbf x_j||$, and there are  $(n-1)n/2$ unique positive distance pairs. Denote the  $(n-1)n/2$   distance pairs $\mathbf d_s=(d_{s,1},...d_{s,(n-1)n/2})^T$ in an increasing order with the subscript $s$ meaning `sorted'. Here we do not need to consider the case when $d_{i,j}=0$, as $\mathbf u_{i,j}=\mathbf 0$, leading to zero contribution to the velocity. Thus the model in (\ref{equ:interaction}) can be reduced to exclude the interaction between particle at zero distance. In reality, two particles at  the same position are impractical, as there typically exists a repulsive force when two particles get very close. Hence, we can reduce the $n^2$ distance pairs $d_{i,j}$ for $i=1,...,n$ and $j=1,...,n$, to $(n-1)n/2$ unique positive  terms  $d_{s,i}$ in an increasing  order, for $i=1,...,(n-1)n/2$. 

Denote the $(n-1)n/2\times (n-1)n/2$ correlation matrix of the kernel  $\bm \phi=(\phi(d_{s,1}),...,\phi(d_{s, (n-1)n/2}))^T$  by $\mathbf R_s$ and $\mathbf U_s $ is $nD\times  (n-1)n/2$ sparse matrix with $n-1$ nonzero terms on each row, where the nonzero entries of the $i$th particle correspond to the distance pairs in the $\mathbf R_s$. Denote the nugget parameter $\eta=\sigma^2_0/\sigma^2$.  After marginalizing out $\phi$, the covariance of velocity observations follows
\begin{align}
(\mathbf {\tilde v} \mid  \gamma, \sigma^2, \eta) \sim \mathcal{MN}\left(\mathbf 0,  \sigma^2 \mathbf {\tilde R}_v \right),
\label{equ:v}
\end{align}
with 
\begin{equation}
\mathbf {\tilde R}_v =(\mathbf U_s \mathbf R_{s} \mathbf U^T_s +\eta \mathbf I_{nD} ).
\label{equ:R_s}
\end{equation}

The conditional distribution of the interaction kernel $\phi(d^*)$ at any  distance $d^*$  follows 
\begin{align}
 (\phi(d^*) \mid  \mathbf {\tilde v}, \gamma, \sigma^2,  \eta)\sim \mathcal N( \hat \phi(d^*),  \sigma^2 K^{*} ),  
 \label{equ:phi_pred_dist}
 \end{align}
where the predictive mean and variance follow
\begin{align}
 \hat \phi(d^*)&= \mathbf r^T(d^*)\mathbf U^T_s  \mathbf {\tilde R}_v ^{-1} \mathbf {\tilde v},  \label{equ:phi_pred_mean}\\
 \sigma^2 K^{*}&=\sigma^2\left( K(d^*,d^*)- \mathbf r^T(d^*)\mathbf U^T_s  \mathbf {\tilde R}_v ^{-1} \mathbf U_s \mathbf r(d^*) \right), 
\end{align}
with $\mathbf r(d^*)=(K(d^*,d_{s,1}),..., K(d^*,d_{s,n(n-1)/2}))^T$. After obtaining the estimated interaction kernel, one can use it to forecast trajectories of particles and understand the physical mechanism of flocking behaviors.

Our primary task is to efficiently compute the predictive distribution of interaction kernel in (\ref{equ:phi_pred_dist}), where the most computationally expensive terms in the predictive mean and variance is $\mathbf {\tilde R}_v ^{-1} \mathbf {\tilde v}$ and $\mathbf {\tilde R}_v ^{-1} \mathbf U_s \mathbf r(d^*)$. Note that the $\mathbf U_s $ is a sparse matrix with $n(n-1)d$ nonzero terms and the inverse covariance matrix $\mathbf R^{-1}_{s}$ is a tri-diagonal matrix. However, directly applying the CG algorithm is still computationally challenging, as neither $\mathbf {\tilde R}_v$ nor $\mathbf {\tilde R}_v ^{-1}$ is sparse. To solve this problem, we extend a step in the Kalman filter to efficiently compute the matrix-vector  multiplication  with the use of sparsity induced by the choice of covariance matrix.  Each step of the CG iteration in the new algorithm  only costs  $\mathcal O(nDT)$ operations for computing a system of $n$ particles and $D$ dimensions with $T$ CG iteration steps. For most systems we explored, we found a few hundred iterations in the CG algorithm achieve high accuracy. The substantial reduction of the computational cost allows us to use more observations to improve the predictive accuracy. We term this approach the sparse conjugate gradient algorithm for Gaussian processes (sparse CG-GP). The algorithm for the scenario with $M$ simulations, each containing $L$ time frames of $n$ particles in a $D$ dimensional space,  is discussed in Appendix \ref{subsec:algorithm_particles}.

\begin{figure}[t] 
\centering
  \begin{tabular}{cc}
  \hspace{-.25in}  
    \includegraphics[width=.5\textwidth]{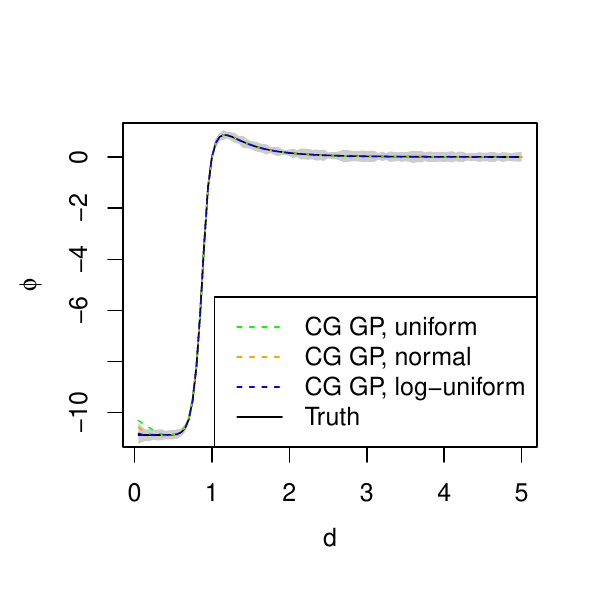}
 \hspace{-.3in}   
 \includegraphics[width=.5\textwidth]{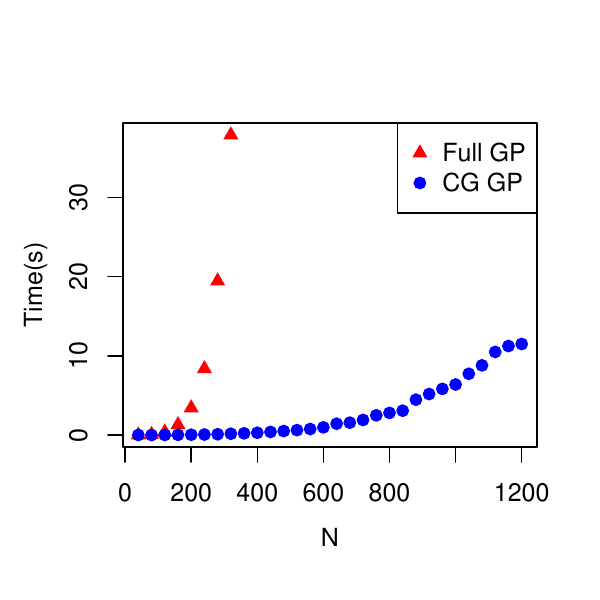}  
\hspace{-.25in}        

        \vspace{-.2in}
  \end{tabular}
   \caption{Estimation of particle interaction kernel by the truncated Lennard-Jones potential in \cite{lu2019nonparametric} with $D=2$. The left panel shows the true interaction kernel (black curve) and estimated kernel functions (colored curves), based on three different ways of initial positions of  $n=200$ particles. The computational time of the full GP in \cite{feng2021data} and the sparse CG-GP approach is given in the right panel. Here the most computational intensive part of the full GP model is in constructing the correlation matrix $\mathbf R_s$ of $\phi$ for $ n(n-1)/2$ distance pairs in Equation (\ref{equ:R_s}), which scales as $\mathcal O(n^4D)$.   }
\label{fig:truncated_LJ}
\end{figure}

The comparison of the computational cost between the full GP model and the proposed sparse CG-GP method is shown in the right panel in Figure \ref{fig:truncated_LJ}. The most computational expensive part of the full GP model is on constructing the $n(n-1)/2\times n(n-1)/2$ correlation matrix $\mathbf R_s$ of $\phi$ for $ n(n-1)/2$ distance pairs. The sparse CG-GP algorithm is much faster as we do not need to construct this covariance matrix; instead we only need to efficiently compute matrix multiplication  by  utilizing  the sparse structure of the inverse of $\mathbf R_s$ (Appendix \ref{subsec:algorithm_particles}). Note  the GP model with an exponential covariance naturally induces a sparse inverse covariance matrix that can be used for faster computation, which is different from imposing a sparse covariance structure for approximation.

In the left panel in Figure  \ref{fig:truncated_LJ},  we show the predictive mean and uncertainty assessment by the sparse CG-GP method for three different designs for sampling the initial positions of particles. From the first to the third designs, the initial value of each coordinate of the particle is sampled independently from a uniform distribution $\mathcal U[a_1,b_1]$,  normal distribution $\mathcal N(a_2,b_2)$, and log uniform (reciprocal) distribution  $\mathcal{LU}[\log(a_3),\log(b_3)]$, respectively. 

For experiments with the interaction kernel being the truncated Lennard-Jones potential  given in Appendix \ref{subsec:algorithm_particles}, we use $a_1=0$, $b_1=5$,  $a_2=0$, $b_2=5$,  $a_3=10^{-3}$ and $b_3=5$ for three designs of initial positions. 
Compared with the first design, the second design of initial positions, which was assumed in \cite{lu2019nonparametric}, has a larger probability mass of distributions near 0.  In the third design, the distributions of the distance between particle pairs are monotonically decreasing, with more probability mass near 0 than those in the first two designs. 
In all cases shown in Figure  \ref{fig:truncated_LJ}, we assume $M=1$, $L=1$ and the noise variance is set to be $\sigma_0=10^{-3}$ in the simulation. For demonstration purposes, the range and nugget parameters are fixed to be $\gamma=5$ and $\eta=10^{-5}$ respectively, when computing the predictive distribution of $\phi$. The estimation of the interaction kernel on large distances is accurate for all different designs, whereas the estimation of the interaction kernel at small distances is not satisfying for the first two designs. When particles are initialized from the third design (log-uniform), the accuracy is better, as there are more particles near each other, providing more information about the particles at small values. This result is intuitive, as the small distance pairs have relatively small contributions to the velocity based on Equation (\ref{equ:interaction}), and we need more particles close to each other to estimate the interaction kernel function at small distances.

The numerical comparison between different designs  allows us to better understand the learning efficiency in different scenarios, which can be used to design experiments. Because of the large improvement of computational scalability compared to previous studies \cite{feng2021data,lu2019nonparametric}, we can accurately estimate interaction kernels based on more particles and longer trajectories.

\subsection{Numerical results}

\begin{figure*}[t] 
\centering 
    \includegraphics[width=.95\textwidth]{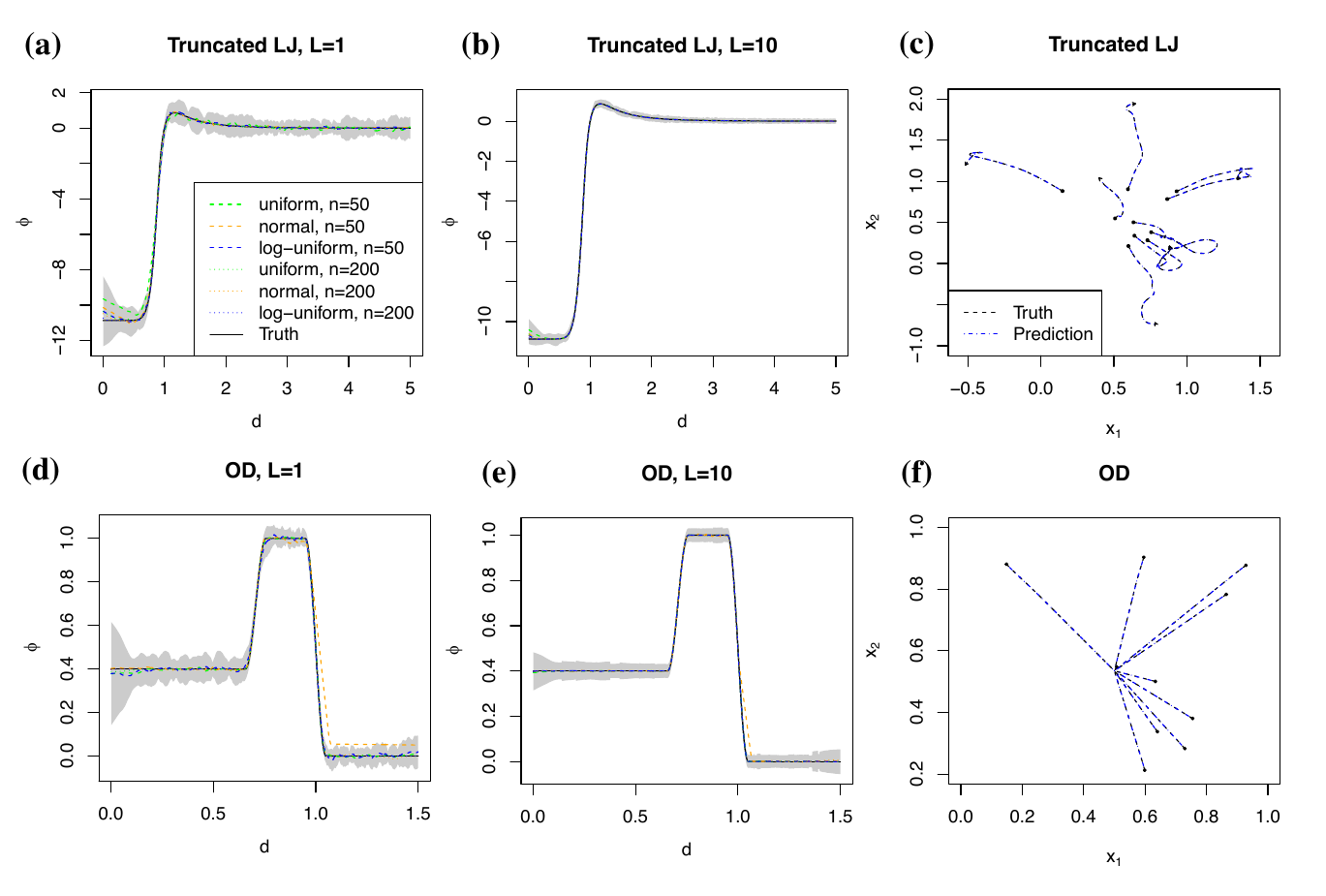}
  \vspace{-.25in}
   \caption{Estimation of interaction function based on the sparse CG-GP method, and trajectory forecast for the truncated LJ and OD simulation. In panel (a), the colored curves are estimated interactions for different initial positions and particle sizes, all based on trajectories using only $L=1$ time step, whereas the black curve is the truncated LJ used in the simulation. The colored curves in panel (b) are the same as those in panel (a), but based on trajectories in $L=10$ time steps. The panels (d) and (e) are the same as panels (a) and (b), respectively, but the simulated data are generated by the OD interaction kernel. The shared areas are the $95\%$ predictive interval. In panel (c),  we graph the simulated trajectories of 10 out of 50 particles $L=200$ time steps, and the trajectory forecast based on estimated interaction function and initial positions. The arrow indicates the direction of velocities of particles at the last time step. Panel (f) is the same as panel (c), but for the OD interaction kernel.   } 
\label{fig:phi_x_pred}
\end{figure*}

Here we discuss two scenarios, where the interaction between particles follow the truncated Lennard-Jones (LJ) and opinion dynamics (OD) kernel functions. The LJ potential is widely used in MD simulations of interacting molecules \cite{rapaport2004art}. First-order systems of form \eqref{equ:interaction} have also been successfully applied  in modeling opinion dynamics in social networks (see the survey \cite{motsch2014heterophilious} and references therein). The interaction function $\phi$ models how the opinions of pairs of people influence each other. In our numerical example, we  consider 
 heterophilious opinion interactions: each agent is more influenced by its neighbors slightly further away from its closest neighbors.  As time evolves, the opinions of agents merge into clusters, with the number of clusters significantly smaller than the number of agents. This phenomenon was studied in  \cite{motsch2014heterophilious} that heterophilious dynamics enhances consensus,  contradicting the intuition that would suggest that the tendency to bond more with those who are different rather than with those who are similar would break connections and prevent clusters of consensus.

The  details of the interaction functions are given in Appendix \ref{subsec:interaction_functions}. For each interaction, we test our method based on 12 configurations of 2 particle sizes ($n=50$ and $n=200$), 2 time lengths ($L=1$ and $L=10$), and 3  designs of initial positions (uniform, normal and log-uniform). The computational scalability of the sparse CG algorithm allows us to efficiently  compute the predictions in most of these experimental settings within a few seconds. For each configuration, we repeat the experiments 10 times to average  the effects of randomness in the initial positions of particles. The root of the mean squared error in predicting the interaction kernels by averaging these 10 experiments of each configuration is given in Appendix  \ref{subsec:numeric_comparison_Appendix}.

In Figure \ref{fig:phi_x_pred}, we show the estimation of  interactions kernels and forecasts of particle trajectories with  different designs, particle sizes and time points. The sparse CG-GP method is relatively accurate for almost all scenarios. 
Among different initial positions, the estimation of trajectories for LJ interaction is the most accurate when the initial positions of the particles are sampled by the log-uniform distribution. This is because there are more small distances between particles when the initial positions follow a log-uniform distribution, providing more data to estimate the interaction kernel at small distances. Furthermore, when we have more particles or observations at larger time intervals, the estimation of the interaction kernel from all designs becomes more accurate in terms of the normalized root mean squared error with the detailed comparison given in Appendix \ref{subsec:numeric_comparison_Appendix}.

In panel (c) and panel (f) of Figure \ref{fig:phi_x_pred}, we plot the trajectory forecast of $10$ particles over $200$ time points for the truncated LJ kernel and OD kernel, respectively. In both simulation scenarios, interaction kernels  are estimated based on trajectories of $n=50$ particles across $L=20$ time steps  with initial positions sampled from the log-uniform design. The trajectories of only  $10$ particles out of $50$ particles are shown for better visualization. 
For trajectories simulated by the truncated LJ, some particles can move very close, since the repulsive force between two particles becomes smaller as the force is proportional to the distance from Equation (\ref{equ:interaction}), and the  truncation of kernel substantially reduces the repulsive force when particles move close. For the OD simulation, the particles move toward a cluster, as expected, since the particles always have attractive forces between each other.
The forecast trajectories are close to the hold-out truth, indicating the high accuracy of our approach.

Compared with the results shown in previous studies \cite{lu2019nonparametric,feng2021data},  estimating the interaction kernels and forecasting trajectories both look more accurate.  The large computational reduction by the sparse CG-GP algorithm shown in Figure \ref{fig:truncated_LJ} permits the use of longer trajectories from more particles to estimate the interaction kernel, which improves the predictive accuracy. Here particle has interactions with all other particles in our simulation, making the number of distance pairs large. Yet   we are able to estimate the interaction kernel and forecast the trajectories of particles within only tens of seconds in a desktop for the most time consuming scenario we considered. Since the particles typically have very small or no interaction when the distances between them are large,   approximation can be made by enforcing interactions between particles within the specified radius, for further reducing the computational cost.

\section{Concluding remarks}
\label{sec:conlusion}

We have introduced scalable marginalization of latent variables for correlated data. We first introduce GP models and reviewed the SDE representation of GPs with Mat{\'e}rn covariance and one-dimensional input. Kalman filter and RTS smoother were  introduced as a scalable marginalization way to compute the likelihood function and predictive distribution, which  reduces the computational complexity of GP with Mat{\'e}rn covariance for 1D input from $\mathcal O(N^3)$ to $\mathcal O(N)$ operations without approximation, where $N$ is the number of observations. Recent efforts on extending scalable computation from 1D input to multi-dimensional input are discussed. In particular, we developed a new scalable algorithm for predicting particle interaction kernel and forecast trajectories  of particles. The achievement is through the sparse representation of GPs in modeling interaction kernel, and then efficient computation for matrix multiplication by modifying the Kalman filter algorithm. An iterative algorithm based on CG can then be applied, which  reduces the computational complexity.   

There are a wide range of future topics relevant to this study. First, various models of spatio-temporal data can be written as random factor models in (\ref{equ:model_latent_factor}) with latent factors modeled as Gaussian processes for temporal inputs. It is of interest to utilize the computational advantage of the dynamic linear models of factor processes, extending the computational tools by relaxing the independence between  prior factor processes in Assumption \ref{assump:1} or incorporating the Toeplitz covariance structure for stationary temporal processes. 
Second, for estimating systems of particle interactions, we can further reduce computation by only considering interactions within a radius between particles. Third, a comprehensively study the experimental design, initialization, and parameter estimation in
will be helpful for estimating latent interaction functions that  can be unidentifiable or weakly identifiable in certain scenarios. Furthermore, velocity directions and angle order parameters are essential for understanding the mechanism of active nematics and cell migration, which can motivate more complex models of interactions.   
Finally, the sparse CG algorithm developed in this work is of interest to reducing the computational complexity of GP models with multi-dimensional input and general designs.

\section*{Acknowledgement}
The work is  partially supported by the National Institutes of Health under Award No. R01DK130067. Gu and Liu acknowledge the partial support from National Science Foundation (NSF) under Award No. DMS-2053423. Fang acknowledges the support from the UCSB academic senate faculty research grants program. Tang is partially supported by   Regents Junior Faculty fellowship, Faculty Early Career Acceleration grant, Hellman Family Faculty Fellowship sponsored by UCSB and the NSF under Award No. DMS-2111303. The authors thank the editor and two referees for their comments that substantially improved the article.

\section*{Appendix}

\subsection{Closed-form expressions  of state space representation of GP having Mat{\'e}rn covariance with $\nu=5/2$ }
\label{sec:close_formed_state_space}

Denote $\lambda=\frac{\sqrt{5}}{\gamma}$, 
$ d_i=|x_i-x_{i-1}|$.  For $i=2,...,N$, $\mathbf G_i $ and $\mathbf W_i$ in (\ref{equ:ctdlm}) have the  expressions below: 

\[\mathbf G_i = \frac{e^{-\lambda  d_i}}{2} \begin{pmatrix}
 \lambda^2  d_i^2+2\lambda d_i+2&2(\lambda  d_i^2+ d_i)  & d_i^2\\ 
 -\lambda^3 d_i^2&-2(\lambda^2  d_i^2-\lambda  d_i-1)  &2 d_i-\lambda  d_i^2 \\ 
\lambda^4 d_i^2-2\lambda^3 d_i& 2(\lambda^3 d_i^2-3\lambda^2 d_i) &\lambda^2 d_i^2-4\lambda  d_i +2
\end{pmatrix}, \]
\[\mathbf W_i =\frac{4\sigma^2\lambda^5}{3}\begin{pmatrix}
W_{1,1}(x_i)  &W_{1,2}(x_i)   &W_{1,3}(x_i)   \\ 
W_{2,1}(x_i)  &W_{2,2}(x_i)  &W_{2,3}(x_i)   \\ 
W_{3,1}(x_i)  &W_{3,2}(x_i)  &W_{3,3}(x_i)
\end{pmatrix},  \]
 with 
\begin{align*}
W_{1,1}(x_i)&=\frac{e^{-2\lambda  d_i} (3+6\lambda d_i+6\lambda^2 d^2_i+4\lambda^3 d^3_i+2\lambda^4 d^4_i)-3 }{-4\lambda^5}, \\
W_{1,2}(x_i)&=W_{2,1}(x_i)=\frac{e^{-2\lambda  d_i}d_i^4}{2}, \\
W_{1,3}(x_i)&=W_{3,1}(x_i)=\frac{e^{-2\lambda  d_i}(1+2\lambda  d_i +2\lambda^2  d^2_i+4\lambda^3 d^3_i-2\lambda^4 d^4_i )-1   }{4\lambda^3}, \\
W_{2,2}(x_i)&=\frac{e^{-2\lambda  d_i}(1+2\lambda  d_i +2\lambda^2  d^2_i-4\lambda^3 d^3_i+2\lambda^4 d^4_i )-1   }{-4\lambda^3}, \\
W_{2,3}(x_i)&= W_{3,2}(x_i)=\frac{e^{-2\lambda  d_i} d_i^2(4-4\lambda d_i+\lambda^2 d^2_i) }{2}, \\
W_{3,3}(x_i)&= \frac{e^{-2\lambda  d_i}(-3+10\lambda^2 d^2_i-22\lambda^2 d^2_i+12\lambda^3 d^3_i-2\lambda^4 d^4_i)+3  }{4\lambda},
\end{align*}
and the stationary covariance of $\bm \theta_i$, $i=1,...,N$, is
\[\mathbf W_1= \begin{pmatrix}
 \sigma^2&0  &-\sigma^2\lambda^2/3 \\ 
 0&\sigma^2\lambda^2/3  &0 \\ 
 -\sigma^2\lambda^2/3 &0 &\sigma^2 \lambda^4 
\end{pmatrix}, \]

The joint distribution of latent states follows $\left(\bm \theta^T(x_1),...,\bm \theta^T(x_N) \right)^T\sim \mathcal{MN}(\mathbf 0, \bm \Lambda^{-1} )$, where the $\bm \Lambda$ is a symmetric block tri-diagonal matrix with the $i$th
diagonal block being $\mathbf W^{-1}_{i}+\mathbf G^T_{i} \mathbf W^{-1}_{i+1} \mathbf G_{i} $ for $i=1,...,{N}-1$, and the $N$th diagonal block being $\mathbf W_{{N}}^{-1}$.  The primary upper off-diagonal block of $\bm \Lambda$ is  $-\mathbf G^T_{i} \mathbf W^{-1}_{i}$, and the primary upper off-diagonal block is  $-\mathbf W^{-1}_{i}\mathbf G_{i} $, for $i=2,...,N$.

Suppose $x_i< x^* < x_{i+1}$. Let $ d_i^*= |x_i^*-x_{i}|$ and $ d_{i+1}^*= |x_{i+1} - x_i^*|$. The ``*" terms $\mathbf G_{i}^*$ and $\mathbf W_{i}^*$ can be computed by replacing  $d_i$ in $\mathbf G_{i}$ and $\mathbf W_{i}$  by $d_i^*$, whereas the  ``*" terms $\mathbf G_{i+1}^*$ and $\mathbf W_{i+1}^*$ can be computed by replacing the $d_i$ in $\mathbf G_{i}$ and $\mathbf W_{i}$  by  $d_{i+1}^*$. Furthermore, $\tilde{\mathbf W}^*_{i+1} = \mathbf W_{i+1}^* + \mathbf G_{i+1}^* \mathbf W_{i}^* (\mathbf G_{i+1}^*)^T$.

\subsection{The sparse CG-GP algorithm for estimating  interaction kernels} 
\label{subsec:algorithm_particles}

Here we discuss the details of computing the predictive mean and variance in (\ref{equ:phi_pred_dist}). The $N$-vector of velocity observations is denoted as $\mathbf {\tilde v}$, where the total number of observations is defined by $N=nDML$. 
To compute the predictive mean and variance, the most computational challenging part is  to compute $N$-vector $\mathbf z=(\mathbf U_s \mathbf R_{s} \mathbf U^T_s +\eta \mathbf I_{N} )^{-1}\mathbf {\tilde v}$. Here  $\mathbf R_s$ and $ \mathbf U_s$  are $\tilde N\times \tilde N$ and $N\times \tilde N$, respectively, where $\tilde N=n(n-1)ML/2$  is the number of non-zero unique distance pairs. Note that both $\mathbf U_s$ and $\mathbf R^{-1}_{s}$ are sparse.  Instead of directly computing the matrix inversion and the matrix-vector multiplication, we  utilize the sparsity structure to accelerate the computation in the sparse CG-GP algorithm. 
In the iteration, 
we need to efficiently compute
\begin{equation}
    \tilde {\mathbf z}= (\mathbf U_s \mathbf R_{s} \mathbf U^T_s +\eta \mathbf I_{N} ) \mathbf z,
    \label{equ:CG_R_z}
\end{equation}
for any real-valued N-vector $\mathbf z$.

We have four steps to compute the quantity in (\ref{equ:CG_R_z}) efficiently. Denote $x_{i,j,m}[t_l]$ the $j$th spatial coordinate of particle $i$ at time $t_l$, in the $m$th simulation, for $i=1,...,n$, $j=1,...,D$,   $l=1,...,L$ and $m=1,...,M$. In the following, we use  $\mathbf x_{\cdot,\cdot,m}[\cdot]$ to denote a vector of all positions in the  $m$th simulation and vice versa. Furthermore, we use $z[k]$ to mean the $k$th entry of any vector $\mathbf z$,  $\mathbf A[k,.]$ and   $\mathbf  A[.,k]$ to mean the $k$th row vector and $k$th column vector of any matrix $\mathbf A$, respectively.  The rank of a particle with position $\mathbf x_{i,.,m}[t_l]$ is defined to be $P= (m-1)Ln+(l-1)n+i$.

First, we reduce the $N \times \tilde N $ sparse matrix $\mathbf U_s$ of distance difference pairs to  an $ N \times n$ matrix $\mathbf U_{re}$, where `re' means reduced, with the $((m-1)LnD+(l-1)nD+(j-1)n+  i_1,i_2)$th entry of $\mathbf U_{re}$ being  $(x_{i_1,j,m}[t_l]-x_{i_2,j,m}[t_l])$, for any $|i_1-i_2|=1,...,n-1$, $i_1\leq n$ and $i_2\leq n$. Furthermore, we create a $\tilde N\times 2$ matrix $\mathbf P_r$ in which the $h$th row records the rank of a distance pair is the $h$th largest in the zero-excluded sorted distance pairs $\mathbf d_s$, where   $P_r[h, 1]$ and  $P_r[h, 2]$ are the rank of rows of these distances in the matrix $\mathbf d_{mat}$, where the $j$th column records the unordered distance pairs of the $j$th particle for $j=1,...,n$. We further assume $P_r[h, 1]>P_r[h, 2]$.  

For any $N$-vector $ \mathbf z$, the $k$th entry of $\mathbf U^T_s  \mathbf z$ can be written as 
\begin{multline}
(\mathbf U^T_s \mathbf z)[k]=\sum_{j_k=0}^{D-1}\mathbf U_{re}\bigg[P_r[k,1]+c_{j_k},P_r[k,2]-(m-1)nL-(l-1)n\bigg]\bigg(z[P_r[k,1]+c_{j_k}]-z[P_r[k,2]+c_{j_k}]\bigg), 
\label{equ:U_t_y}
\end{multline}
where $c_{j_k}=(D-1)(m-1)nL+(D-1)(l-1)n+nj_k$
for $k=1,...,\tilde N$, if the $k$th largest entry of distance pair is from time frame $l$ in the $m$th simulation. The output is denoted as an $\tilde N$ vector $\mathbf{g}_{1}$, i.e. $\mathbf{g}_{1}=(\mathbf U^T_s \mathbf z)$.

Second, since the exponential kernel is used,  $\mathbf R^{-1}_s$ is a tri-diagonal matrix \cite{Gu2018robustness}.  We modify  a Kalman filter step to efficiently compute the product of an upper bi-diagonal $\mathbf{g}_{2}=\mathbf L^T_s \mathbf{g}_{1}$, where
 $\mathbf L_s $ is the factor of the Cholesky decomposition $\mathbf R_s=\mathbf L_s \mathbf L^T_s$. 
Denote the Cholesky decomposition of the inverse covariance the factor $\mathbf R^{-1}_s=\mathbf {\tilde L}_s \mathbf {\tilde L}_s^T$, where $ \mathbf {\tilde L}_s$ can be written as the lower bi-diagonal matrix below:
  \begin{equation}
  \mathbf {\tilde L}_s=\begin{pmatrix} 
\frac{1}{\sqrt{1-\rho_1^2}} &  &    &  \\ 
\frac{-\rho_1}{\sqrt{1-\rho_1^2}} & \frac{1}{\sqrt{1-\rho_2^2}} &  &    \\
  & \frac{-\rho_2}{\sqrt{1-\rho_2^2}} \,  \ddots&  &   & \\

  & \quad \ddots & \frac{1}{\sqrt{1-\rho_{\tilde N-1}^2}} &   \\
  & &  \frac{-\rho_{\tilde N-1}}{\sqrt{1-\rho_{\tilde N-1}^2}}  & 1  \\
\end{pmatrix},
\label{equ:tilde_L_s}
\end{equation}
where $\rho_k=\exp(-(d_{s,k+1}-d_{s,k})/\gamma)$ for $k=1,...,\tilde N-1$.
We modify the Thomas  algorithm \cite{thomas1949elliptic} to solve  $\mathbf{g}_{2}$ from equation $(\mathbf L^T_s)^{-1} \mathbf{g}_{2}= \mathbf{g}_{1} $. Here $(\mathbf L^T_s)^{-1}$ is an upper bi-diagonal matrix with explicit form
  \begin{equation}
  (\mathbf L^T_s)^{-1}=\begin{pmatrix} 
1  & \frac{-\rho_1}{\sqrt{1-\rho_1^2}} &      \\ 
 & \frac{1}{\sqrt{1-\rho_1^2}}  &    \\
  && \ddots&   \ddots  &    \\ 
 & &  & \frac{1}{\sqrt{1-\rho_{\tilde N-2}^2}} & \frac{-\rho_{\tilde N-1}}{\sqrt{1-\rho_{\tilde N-1}^2}}  \\
&  & &   &  \frac{1}{\sqrt{1-\rho_{\tilde N-1}^2}}  \\
\end{pmatrix}.
\label{equ:L_inv_s}
\end{equation}
Here only up to 2 entries in each row of $(\mathbf L^T_s)^{-1}$  are nonzero. Using a backward solver, the $\mathbf{g}_{2}$ can be obtained by the iteration below:
\begin{align}
   \mathbf{g}_{2}[\tilde N] &= \mathbf{g}_{1}[\tilde N] \sqrt{1-\rho_{\tilde N-1}^2}, \label{equ:L_inv_t_Thomas_1}\\
     \mathbf{g}_{2}[k] &=   \sqrt{1-\rho_{k-1}^2} \mathbf{g}_{1}[k] + \frac{\rho_k   \mathbf{g}_{2}[k+1] \sqrt{1-\rho_{k-1}^2} } {\sqrt{1-\rho_{k}^2}}, \quad \label{equ:L_inv_t_Thomas_2}
\end{align}
 for $k=\tilde N-1,...,2,1$. 
Note that the Thomas algorithm is not stable in general, but here the stability issue is greatly improved, as the matrix in the system is bi-diagonal instead of tri-diagonal. 

Third, we compute $\mathbf{g}_{3}=\mathbf L_s \mathbf{g}_{2}$ by solving $\mathbf L^{-1}_s \mathbf{g}_{3}= \mathbf{g}_{2} $:
\begin{align}
   \mathbf{g}_{3}[1] &= \mathbf{g}_{2}[1] , \label{equ:L_inv_Thomas_1}\\
     \mathbf{g}_{3}[k] &=  \sqrt{1-\rho_{k-1}^2} \mathbf{g}_{2}[k] + \rho_{k-1}   \mathbf{g}_{3}[k-1], \label{equ:L_inv_Thomas_2}
\end{align}
for $k=2,....,\tilde N-1$.

Finally, we denote a $MLn\times n$ matrix $\mathbf P_c$. $\mathbf P_c$ is initialized as a zero matrix. And then for $r_c=1,...,MLn$, row $r_c$ of $\mathbf P_c$ stores the ranks of distances between the $i$th  particle and other $n-1$ particles in $\mathbf d_s$. For instance, at the $l$th time step in the $m$th simulation, particle $i$ has $n-1$ non-zero distances $||\mathbf x_1-\mathbf x_i ||,...,||\mathbf x_{i-1}-\mathbf x_i ||,||\mathbf x_{i+1}-\mathbf x_i ||,...,||\mathbf x_n-\mathbf x_i ||$  with ranks $h_1, ...., h_{i-1},h_{i+1},... h_{n}$ in $\mathbf d_s$. Then the  $((m-1)Ln+(l-1)n+i)$th  row of $\mathbf P_c$ is filled with $(h_1,...,h_{i-1},h_{i+1},...,h_n)$. 

 Given any $\tilde N$-vector $\mathbf g_{3}$, the $k$th entry of $\mathbf U_s \mathbf g_3$ can be written as 
\begin{equation}
(\mathbf U_s \mathbf g_{3} )[k]=\mathbf U_{re}[k,.] \mathbf g_{3}[\mathbf P_c[k',.]^T], 
\label{equ:U_z}
\end{equation}
assuming that $k$ satisfies $k=(m-1)LnD +(l-1)nD+jn+i$ and $k'=i+(m-1)Ln+(l-1)n$ for some $m, l, j$ and $i$, and  $k=1,...,N$. 
The output of this step is an $N$ vector $\mathbf{g}_{4}$, with the $k$th entry being $\mathbf{g}_{4}[k]:= (\mathbf U_s  \mathbf g_{3} )[k]$, for $k=1,...,N$.

 We summarize the sparse CG-GP algorithm using the following steps to compute  $ \tilde {\mathbf z}$ in (\ref{equ:CG_R_z}) below. 

\begin{enumerate}
    \item Use equation (\ref{equ:U_t_y}) to compute $\mathbf{g}_{1}[k]= (\mathbf U^T_s  \mathbf z)[k]$, for $k=1,...,\tilde{N}$.
    \item Use equations (\ref{equ:L_inv_t_Thomas_1}) and  (\ref{equ:L_inv_t_Thomas_2}) to solve $\mathbf{g}_{2}$ 
    from $ (\mathbf L_s^{T})^{-1} \mathbf{g}_{2}=\mathbf{g}_{1} $ where  $(\mathbf L_s^{T})^{-1}=(\mathbf L_s^{-1})^T$ with $\mathbf L_s^{-1}$ given in equation (\ref{equ:L_inv_s}). 
    \item Use equations (\ref{equ:L_inv_Thomas_1}) and (\ref{equ:L_inv_Thomas_2}) to solve $\mathbf{g}_{3}$ from  $\mathbf L_s^{-1} \mathbf{g}_{3}=\mathbf{g}_{2} $, where $\mathbf L_s^{-1} $ is given in equation (\ref{equ:L_inv_s}). 
    \item Use equation (\ref{equ:U_z}) to compute  $\mathbf{g}_{4}[k] =(\mathbf U_s  \mathbf{g}_{3})[k] $ and let $ \tilde {\mathbf z}=\mathbf{g}_{4} +\eta \mathbf z$.
\end{enumerate}

\subsection{Interaction kernels  in simulated studies}
\label{subsec:interaction_functions}

Here we give the expressions of the truncated L-J and OD kernels of particle interaction in \cite{lu2019nonparametric,feng2021data}. 
  The truncated LJ kernel is given by 
\begin{equation*}
\phi_{LJ}(d) = \left\{
\begin{array}{ll}
  c_2 \exp(-c_1 d^{12}),  & d \in[0, 0.95], \\
  \frac{8(d^{-4} - d^{-10})}{3},  & d \in (0.95, \infty),
\end{array}
\right.
\end{equation*}
where 
\begin{align*}
 c_1 = -\frac{1}{12} \frac{c_4}{c_3 (0.95)^{11}} \mbox{ and } c_2 = c_3 \exp(c_1(0.95)^{12}), 
    \end{align*}
with $c_3 = \frac{8}{3}(0.95^{-4} - 0.95^{-10})$ and $c_4 = \frac{8}{3}(10(0.95)^{-11} - 4(0.95)^{-5})$.

The interaction kernel of OD is defined as
\begin{equation*}
\phi_{OD}(d) = \left\{
\begin{array}{ll}
  0.4,  & d\in [0, c_5), \\
  -0.3\cos(10\pi(d-c_5)) +0.7, & d \in [c_5, c_6),\\
  1, &  d \in [c_6 \leq d < 0.95),\\
  0.5\cos(10\pi(d-0.95)) +0.5,  & d\in [0.95, 1.05),\\
  0, & d \in [1.05, \infty), 
\end{array}
\right.
\end{equation*}
where $c_5=\frac{1}{\sqrt{2}}-0.05$ and $c_6=\frac{1}{\sqrt{2}}+0.05$.

\subsection{Further numerical results on estimating interaction kernels}
\label{subsec:numeric_comparison_Appendix}
We outline the numerical results of estimating  the interaction functions at $N^*_1=1000$ equally spaced distance pairs at $d\in [0,5]$ and $d\in [0, 1.5]$ for the truncated LJ and OD, respectively. For each configuration,  we repeat the simulation  $N^*_2=10$ times and compute the predictive error in each simulation.  The total number of test points is $N^*=N^*_1N^*_2=10^4$.  For demonstration purposes, we do not add a noise into the simulated data (i.e. $\sigma^2_0=0$). The range and nugget parameters are fixed to be $\gamma=5$ and $\eta=10^{-5}$.  We compute the normalized root of mean squared error (NRMSE) in estimating the interaction kernel function: 
\[\mbox{NRMSE}=\frac{1}{\sigma_{\phi}}\sqrt{ \sum^{N^*}_{i=1}\frac{(\hat \phi(d^*_i)-\phi(d^*_i) )^2 }{N^*} }, \]  
where $\hat \phi(.)$ is the estimated interaction kernel from the velocities and positions of the particles;  $\sigma_{\phi}$ is the standard deviation of the interaction function at test points. 

\begin{table}[h]
\begin{center}
\begin{tabular}{lllll}
  \hline
     Truncated LJ                & n=50   &  n=200  &  n=50 & n=200  \\
     & L=1 &  L=1 & L=10& L=10   \\
  \hline
     Uniform  & $.11$  & $.021$  & $.026$  & $.0051$  \\
    Normal  & $.037$  & $.012$  & $.0090$  & $.0028$  \\
    Log-uniform  & $.043$  & $.0036$  & $.0018$  & $.00091$  \\
    \hline
       OD               & n=50  &  n=200  &  n=50 & n=200   \\
       & L=1 & L=1  & L=10 & L=10  \\
                         \hline
    Uniform   & $.024$  & $.0086$  & $.0031$  & $.0036$  \\
    Normal & $.13$  & $.013$  & $.038$  & $.0064$  \\
    Log-uniform          & $.076$  & $.0045$  & $.0018$  & $.00081$  \\
                         \hline

\end{tabular}
\end{center}
\caption{NRMSE of the sparse CG-GP method for estimating the truncated LJ and OD kernels of particle interaction. }
\label{tab:est_phi}

\end{table}

Table \ref{tab:est_phi} gives the NRMSE of the sparse CG-GP method for the truncated LJ and OD kernels at 12 configurations. Typically the estimation is the most accurate when the initial positions of the particles are sampled from the log-uniform design for a given number of observations and an interaction kernel. This is because  the contributions to the velocities from the kernel function are proportional to the distance of particle in (\ref{equ:interaction}), and small contributions from the interaction kernel at small distance values make the kernel hard to estimate from the trajectory data in general. When  the initial positions of the particles are sampled from the log-uniform design, more particles are close to each other, which provides more information to estimate the interaction kernel at a small distance. 

Furthermore,   the predictive error of the interaction kernel is smaller, when the trajectories with a larger number of particle sizes or at longer time points are used in estimation, as more observations typically improve predictive accuracy.   The sparse CG-GP algorithm reduces the computational cost substantially, which allows more observations  to be used for making predictions. 


\bibliographystyle{plain}
\bibliography{References_2022}

\end{document}